\documentclass[bibyear]{aa}
 \pdfoutput=1
\usepackage{amsmath}
\usepackage{amssymb}
\usepackage{paralist}
\usepackage{natbib}
\usepackage{caption,graphicx,subcaption,mathtools}
\usepackage{graphicx}

\usepackage{txfonts}
\usepackage{tikz}
\usetikzlibrary{shapes,arrows,calc}

\newcommand\diff{{\rm d}}
\newcommand{\R}{{\mathbb R}}

\begin{document}

\title{ADAM: a general method for using various data types in asteroid reconstruction} 

\author{Matti Viikinkoski\inst{1}
\and Mikko Kaasalainen\inst{1}
\and Josef \v{D}urech\inst{2}
}
\institute{Department of Mathematics, Tampere University of Technology,
PO Box 553, 33101 Tampere, Finland
 \and
Astronomical Institute, Faculty of Mathematics and Physics, Charles University in Prague, V Hole\v{s}ovi\v{c}k\'ach 2, 18000 Prague, Czech Republic
}

\date{Received; accepted}
\titlerunning{Disk-resolved data in asteroid modelling}
\authorrunning{M. Viikinkoski et al.}

\abstract{We introduce ADAM, the All-Data Asteroid Modelling algorithm. ADAM is simple and universal since it handles all disk-resolved data types (adaptive optics or other images, interferometry, and range-Doppler radar data) in a uniform manner via the 2D Fourier transform, enabling fast convergence in model optimization. The resolved data can be combined with disk-integrated data (photometry). In the reconstruction process, the difference between each data type is only a few code lines defining the particular generalized projection from 3D onto a 2D image plane. Occultation timings can be included as sparse silhouettes, and thermal infrared data are efficiently handled with an approximate algorithm that is sufficient in practice due to the dominance of the high-contrast (boundary) pixels over the low-contrast (interior) ones. This is of particular importance to the raw ALMA data that can be directly handled by ADAM without having to construct the standard image. We study the reliability of the inversion by using the independent shape supports of function series and control-point surfaces. When other data are lacking, one can carry out fast nonconvex lightcurve-only inversion, but any shape models resulting from it should only be taken as illustrative global-scale ones.} 
\maketitle

\section{Introduction}
Groundbased and other remote-sensing data on asteroids are obtained with a variety of instruments that essentially sample regions on the surface of the target in various ways. These share some common mathematical characteristics of generalized projections \citep{KL06,Ka11,VK14}. The most abundant source of data for asteroid shape and spin reconstruction is disk-integrated photometry, because even datasets sparse in time are often sufficient for modelling \citep{Ka04,DJ06}. Lightcurve-inversion procedures \citep{KT01b} are available at, e.g., the Database of Asteroid Models from Inversion Techniques (DAMIT) site\footnote{\url{http://astro.troja.mff.cuni.cz/projects/asteroids3D}}. Due to the inherently limited information content of the disk-integrated data, the corresponding models are usually most reliably described in convex space (\cite{DK03}, and further discussed below). However, even partially disk-resolved data offer a realistic possibility of more detailed modelling. Previously described approaches for such reconstruction are the SHAPE software \citep{OH02} for radar and lightcurve data, and the KOALA procedure \citep{KV12,CK12} for optical images, occultation timings, and lightcurves.

\subsection{The whole is more than the sum of its parts}
The best way to reconstruct a model of an asteroid is to use all available data. To combine disk-resolved data (adaptive optics or other images, interferometry, and range-Doppler radar data) with disk-integrated data (photometric or infrared lightcurves) and occultation timings (``sparse silhouettes''), we need a general procedure for using any data sources in asteroid modelling. We call this ADAM: All-Data Asteroid Modelling. Concise accounts of the various data types and their modelling aspects are given in \cite{KL06}, \cite{KD13}, and \cite{Du15}. This paper is intended as a technical companion to those reviews.

We present here the ADAM algorithm in a high-level format that includes all the necessary methods and formulae, either written here or given by references to the literature. We discuss and collect here the essential techniques and aspects of a complete inversion procedure capable of handling all the major asteroid data sources and formats. The key point is that complementary data sources can facilitate a good reconstruction even when none of them is sufficient alone. 

The paper is organized as follows. In Sect. 2 we describe the various shape supports we use in the reconstruction; some with the emphasis on global features, some concentrating on local details. This is intimately connected with the reliability estimate of the result, since independent shape representations help to reveal which features are the most probable ones. Sect.\ 3 introduces the Fourier transform method necessary for a simple and universal handling of data sources of disk-resolved type. In Sect.\ 4, we present examples of such types (interferometry, radar, and optical images). The interferometric data from ALMA are of particular interest here.  We also discuss the special case of one-dimensional projections (continuous-wave radar and certain types of interferometry). In Sect.\ 5 we sum up everything in the form of the ADAM algorithm, and conclude in Sect.\ 6. Some basic ADAM functions are listed in an Appendix.

\subsection{ADAM software package}
Using the methods and algorithm described here and in \cite{KT01b}, \cite{Ka11}, and \cite{KV12}, writing an ADAM program from scratch is quite straightforward (for example, convex lightcurve inversion is inherently more complex). We have uploaded free-to-use ADAM code files and functions written in Matlab and C to a toolbox at the DAMIT site. These can be used for writing customized inversion software, and for browsing and understanding the computational methods. The latter, too numerous to be discussed here in detail, include techniques such as the partial derivative chains for gradient-based optimization, ray-tracing procedures, projections and transforms, scattering and luminosity models, GPU acceleration, etc. (Note that we do not offer any user support: the files are presented as is.) 

ADAM is a considerably more general package than the KOALA procedure \citep{KV12,CK12} that is based on extractable image contours. The KOALA contour-fitting principle is necessary for including occultation data, so a full ADAM procedure inherits this function from KOALA. For fitting any pixel images, we recommend the ADAM Fourier-transform functions rather than KOALA.

We take asteroid reconstruction to mean here that the following output parameters are derived form input data: 1) shape (surface) definition, 2) rotational state (period and spin axis direction; possibly also terms for YORP acceleration, precession, or a binary orbit), 3) scattering or other luminosity parameters (often fixed a priori), and 4)  image offset (alignment) and possible other auxiliary or normalization terms. Without loss of generality, we do not discuss each item separately, but mostly take the shape parameters to represent all the free ones since the optimization principle is technically the same for all parameter types. The speed, convergence, and reliability of gradient-based optimization methods are here superior to global methods \citep[such as genetic algorithms or Monte Carlo; see the discussion in][]{KT01b}. We emphasize that the spin parameters, especially the period, usually have numerous local minima, so a dense enough comb of initial values of these is a prerequisite for a good final reconstruction. 

\section{Shape}
Given the diverse shapes of asteroids and the continuing progress in instrument technology, effective methods for shape representation are required for a general reconstruction scheme from observations. In inverse problems it is typically not clear a priori how well a given shape support will perform. In this section we present shape supports and corresponding regularization functions well suited for asteroid-like shapes.   

\subsection{Shape supports}
An important part of shape modelling is the choice of shape representation. Assuming a typical asteroid surface is homeomorphic to the unit sphere, we can consider each coordinate as a function on the sphere, and choosing a suitable basis for functions, expand coordinate functions using this basis. This is straightforward to generalize to multiple bodies such as binaries. Typical such bases are spherical harmonics, spherical wavelets, and spherical splines. Our experiments suggest that parametrizations which expand each coordinate function separately tend to produce suboptimal results since they ignore the geometric dependencies and constraints between coordinates when considering surfaces represented by non-tangled meshes. Thus we have found it useful to consider two well-regulated but conceptually different shape supports in practice: octantoids based on spherical harmonics, and subdivision surfaces.
\subsubsection{Function series}

 An octantoid is a surface given by $p\in\R^3$ that can be parametrized in the form

\begin{equation}
\label{eq:oct} p(\theta,\varphi)=
\left\{\begin{array}{lr}
x(\theta,\varphi)=& e^{a(\theta,\varphi)}\sin\theta \cos\varphi,\\
			    y(\theta,\varphi)=&e^{a(\theta,\varphi)+b(\theta,\varphi)}\sin\theta \sin\varphi,\\
			    z(\theta,\varphi)=&e^{a(\theta,\varphi)+c(\theta,\varphi)}\cos\theta,
                             \end{array}\right.
\end{equation}
where $a$, $b$ and $c$ are conveniently expressed as linear combinations of the (real) spherical harmonic functions $Y^m_l(\theta,\varphi)$, with coefficients $a_{lm}$, $b_{lm}$ and $c_{lm}$, respectively. Note that $(\theta,\varphi)$, $0\le\theta\le\pi$, 
$0\le\varphi<2\pi$, are coordinates on the unit sphere $S^2$ parametrizing the surface but not 
describing any physical directions such as polar coordinates. As usual, the Laplace series for $a,b,c$ are useful for keeping the number of unknowns; i.e., the coefficients of $Y_l^m$, small and the surface smooth. If $b=c=0$, this representation is the usual starlike one with the radius $\exp(a)$, but we have found that even if the target is starlike, the octantoid form allows the capture of detail better, and $b$ and $c$ can be represented with considerably fewer terms than the main function $a$. The number of shape parameters is thus between the $(l_{\rm max}+1)^2$ of the starlike case and $3(l_{\rm max}+1)^2$, when $l_{\rm max}$ is the largest degree of the function series.
The drawback of this representation is its globality: one might want less smoothness regularization in some regions than in others. When more local control is desired (e.g., a feature clearly visible in fly-by images or in radar), the representation (\ref{eq:oct}) may be expanded  with spherical splines or spherical wavelets to provide local detail without affecting the global shape. Depending on the desired level of resolution and the non-starlike irregularity of the surface, the number of free function series coefficients is typically between 50 and 300 from low- to mid-resolution. Function series are seldom useful for high resolution, where one may ultimately want to adjust each vertex separately by defining individual $a_i$, $b_i$, and $c_i$.

\subsubsection{Subdivision control points}
Subdivision surfaces offer local control more than global representations like function series. Beginning with an initial set of vertices and corresponding triangles, called a control mesh, the surface is iteratively refined by adding new vertices and computing new positions for old vertices. The vertex coordinates of the control mesh form the parameter set defining the surface. Each subdivision step smoothes out the surface in a higher level of resolution. Well-behaving subdivision schemes converge to a smooth limit surface.

In this paper we use the Loop subdivision scheme \citep{Lo97}. Considering a vertex $p$ with immediate neighbours $p_0,\ldots,p_{n-1}$, the subdivision method first creates new vertices by splitting each edge:
\begin{equation}
q_i=\frac{3p+3p_i+3p_{i-1}+p_{i+1}}{8},\qquad i=0,\ldots,n-1,
\end{equation}
where the indices should be interpreted as modulo $n$. After the vertex creation step, the position of the vertex $p$ is refined:
\begin{equation}
p'=(1-n\beta)p+\beta\sum p_i.
\end{equation}
The multiplier $\beta$ is usually chosen to be
\begin{equation}
\beta=\frac{1}{n}\left[\frac{5}{8}-\frac{(3+2\cos(2\pi/{n}))^2}{64}\right],
\end{equation}
but other choices are also possible.
The limit surface is continuous; $C^2$ at the ordinary vertices (i.e. vertices that have $6$ neighbours) and $C^1$ at extraordinary vertices. The number of free control points for model rendering is similar to or somewhat lower than the number of function series coefficients (for a comparable level of resolution).

\begin{figure}

        \centering
        \begin{subfigure}[b]{0.3\columnwidth}
                \includegraphics[clip=true,trim=190 300 200 300,scale=0.35]{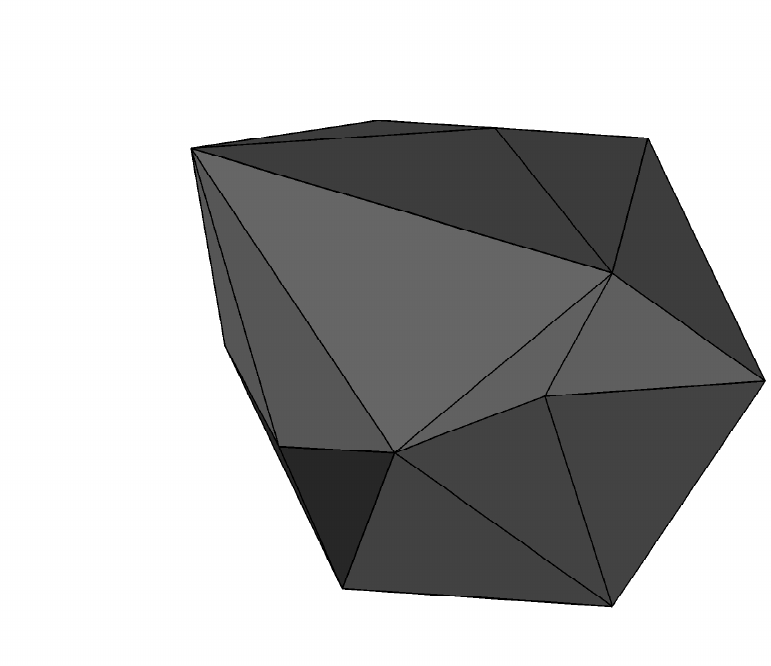}
        \end{subfigure}%
         \begin{subfigure}[b]{0.3\columnwidth}
                \includegraphics[clip=true,trim=190 300 200 300,scale=0.35]{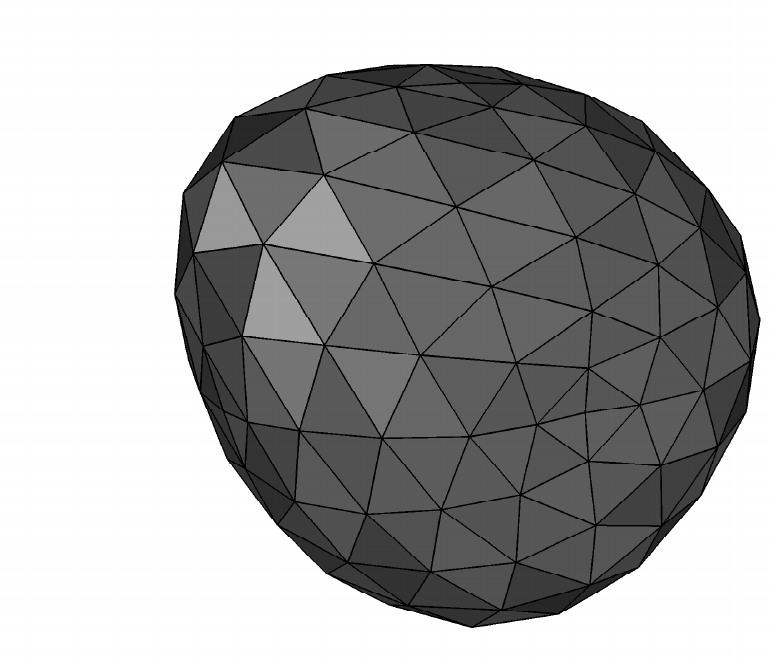}
        \end{subfigure}%
         \begin{subfigure}[b]{0.3\columnwidth}
                \includegraphics[clip=true,trim=190 300 200 300,scale=0.35]{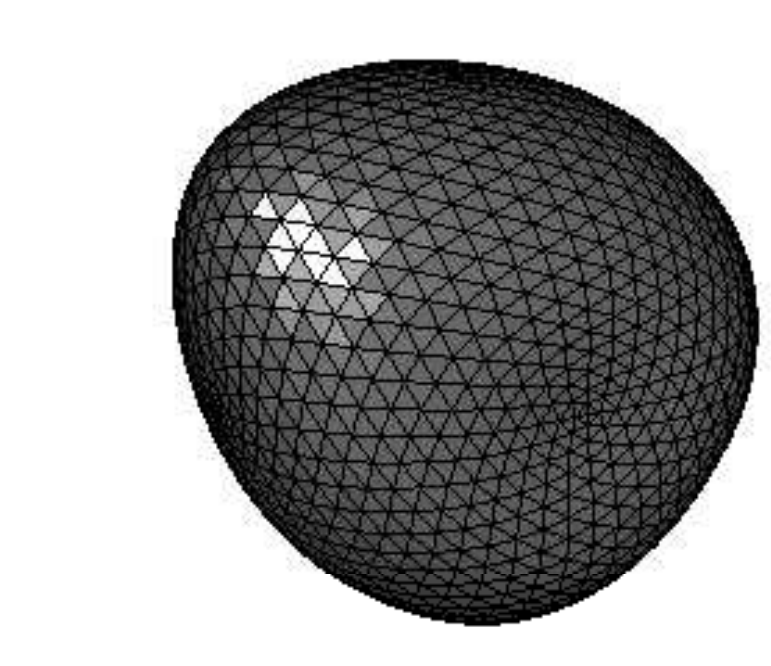}
        \end{subfigure}%
        \caption{Original control mesh (left) with 18 vertices (54 coordinates) as the shape parameter set and 32 facets, after two $\sqrt{3}$-subdivision steps (middle), and after four subdivision steps (right).}
        \label{sqrt3}
        \end{figure}

The main computational aspect with subdivision methods is that the number of facets increases exponentially with the number of divisions. After $n$ subdivision steps, each facet that has been divided has produced $4^n$ subfacets. An alternative scheme to Loop subdivision is the $\sqrt{3}$-subdivision \citep{Ko02}. Instead of splitting the edges, $\sqrt{3}$-scheme subdivides facets by inserting a new vertex to the facet centroid and connecting it to the vertices of the facet (Fig.\ \ref{sqrt3}). 
The main attraction of the $\sqrt{3}$-scheme compared to the Loop subdivision is the slower increase ($3^n$) of facets, while performing similarly in the limit.

In practice, it is usually a good idea to choose the initial control mesh to be an ellipsoid or a scaled convex surface obtained from lightcurve inversion, with a suitable number of vertices for the mesh.  The number of subdivision steps should be chosen carefully: while each subdivision increases resolution and stability by spreading the influence of each parameter to a larger number facets, the computational burden grows exponentially.
Instead of subdividing all the facets, better performance may be obtained with adaptive subdivision, where only facets benefiting from increased resolution are subdivided. However, how to do this automatically during optimization is not obvious. A heuristic inclusion of surface regions to be refined based on a ranking of the improvement of the fit is one possibility \citep[cf.\ the $\chi^2$-sensitivity map of][]{KV12}; visual inspection of the model fit and a graphical user interface can guide in this.

\subsection{Regularization functions}
In inverse problems, finding a feasible regularization method is typically the most delicate part of problem solving. Ideally, both the shape representation and regularization method should be chosen to complement each other. The shape support should be general enough to represent probable shapes, and the regularization should prevent unrealistic or degenerate shapes while, at the same time, reveal the features present in the data.
For octantoids, the choice is remarkably easy. Assuming the basic shape is geometrically starlike, it is intuitively obvious to penalize the deviation from starlikeness. To this effect, we define
\begin{equation}
\eta=\sum_{l,m} l (b_{lm}^2+c_{lm}^2).
\end{equation}
Every starlike surface has a representation for which $\eta=0$, so $\eta$ is a natural quantity to be included in the final $\chi^2$-function to be minimized (Sect.\ 5). The $\chi^2$-sum contains both the goodness-of-fit measure and the regularizing functions that represent prior assumptions and expectations of the solution.

Subdivision surfaces have somewhat different smoothness properties in this regard. It is well known that the Loop subdivision converges to a smooth surface, so each subdivision step will produce a smoother result. However, it is computationally expensive to take a large number of subdivision steps.  Therefore it is advantageous to  combine a few, usually two or three, subdivision steps with  mesh-based regularization methods.

While not strictly necessary, it is convenient to assume that the triangular mesh representing the shape forms a manifold. This assumption makes the checking of shadowing and illumination both conceptually and computationally simpler. Thus it is  imperative to avoid self-intersections, as they  introduce errors to the fitting process. One approach is to explicitly check for intersecting facets and retriangulate if required. However, triangulation and intersection tests are costly, and usually optimization steps leading to self-intersections are suboptimal. A better approach is to prevent self-intersections in the first place. 

Regularization based on dihedral angles penalizes large angles between adjacent facet normals; i.e., the regularization prefers planar regions. We thus want to minimize
\begin{equation}
\gamma_1= \sum_{i,j\in{\mathcal T}} w_{ij}(1-\nu_i \cdot\nu_j),
\end{equation}
where $\mathcal{T}$ are the facets of the mesh, and $\nu_k$ is the unit normal vector corresponding to the facet $k$. The sum is over all those facets $j$ that are adjacent to the facet $i$, and the weights $w_{ij}$ are usually chosen to be unity. As a special case, we may suppress only concave features, obtaining convex regularization \citep{KV12}:
\begin{equation}
\gamma_2=\frac{1}{\sum_j A_j}\sum_{i,j} A_j(1-\nu_i\cdot \nu_j),
\end{equation}
where $A_i$ is the area of the facet $i$ and the sum is over those facets $j$ that are adjacent to the facet $i$ and tilted above its plane.

To prevent degenerate facets and maintain a homogeneous mesh, it is advantageous to inhibit large variations in facet areas:
\begin{equation}
\gamma_3=\sum_i(A_i-\bar{A})^2,
\end{equation}
where $\bar{A}$ is the mean facet area of the polyhedron.

In practice, the regularization functions $\eta$ and $\gamma_2$ are sufficient for octantoid surfaces, while $\gamma_2$  and $\gamma_3$ are useful for the subdivision surfaces. Unrealistically sharp angles can be prevented with $\gamma_1$, but too large a weight will inhibit convergence. 
\indent In addition to geometric considerations, one can use regularization based on physical constraints, such as the requirement for the rotation axis to be close to the largest principal axis of the inertia tensor \citep{Ka11,KV12}.
\subsection{Reliability estimates}
The octantoid representation or the subdivision mesh tend to produce aesthetically pleasing, ``asteroid-like'' surfaces, but it is not initially obvious which surface features of the model are actually present in the data, and which are the side effects of the shape support and the regularization used. 
Conventionally, Markov chain Monte Carlo (MCMC) methods are used to obtain a reliability estimate for the model parameters. However, in our case, modelling  and systematic errors usually dominate \citep{KD06}, rendering the MCMC approach inefficient and inaccurate because the error distribution is not known (it is certainly  not random Gaussian as in standard MCMC). 

Moreover, the posterior distribution of shape parameters from MCMC will not really tell anything about the reliability of the model with respect to data, but only about the distribution of the estimate within the adapted shape support. We have found that this results in an overoptimistic conception of the reliability of the result, simply because the acceptable shape results cannot be probed widely enough using one shape support only -- the Monte Carlo procedure focuses on too small regions of shape variation for both computational and geometric reasons. In addition, the computation of the model fit is time consuming  if the data set and parameter space are large, making MCMC estimation computationally expensive.

\begin{figure}

        \centering
        \begin{subfigure}[b]{0.25\textwidth}
                \includegraphics[clip=true,trim=190 300 200 300,scale=0.35]{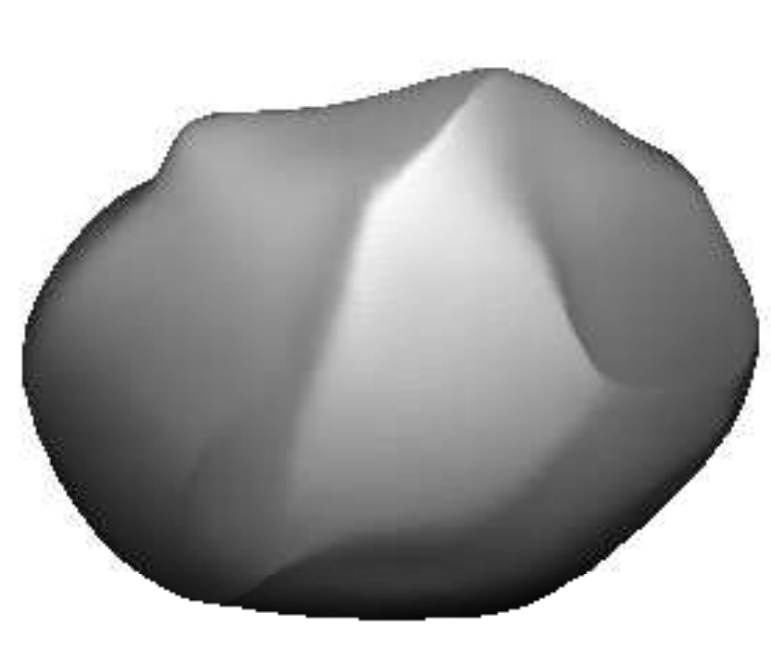}
        \end{subfigure}%
        \begin{subfigure}[b]{0.25\textwidth}
                \includegraphics[clip=true,trim=190 300 200 300,scale=0.35]{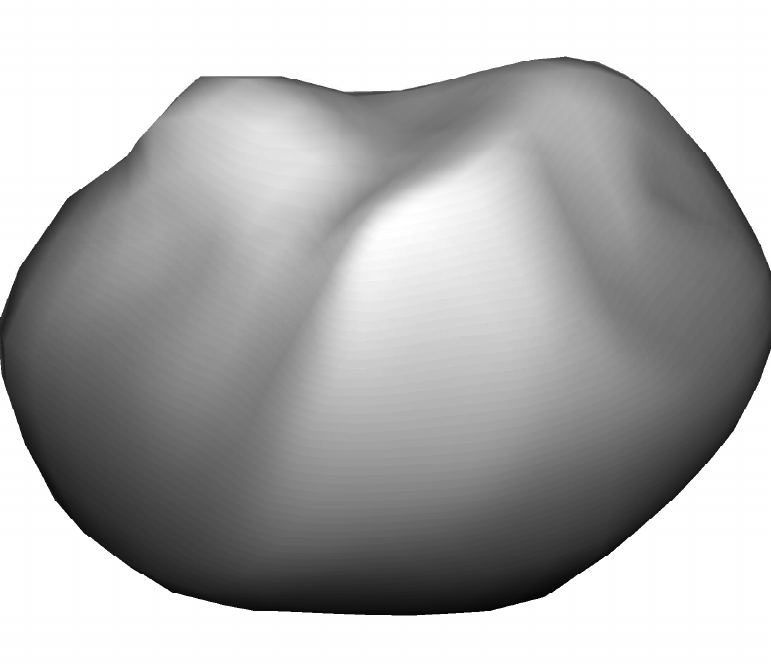}
        \end{subfigure}
        \caption{\label{fig:daphne}Model of asteroid (41)~Daphne from adaptive optics images, reconstructed as a subdivision surface (left) and an octantoid (right).}
        \end{figure}

To circumvent these obstacles, we have found the following approach fast and robust in practice. Any real feature of the model based on the data should also be present if another, independent model type such as shape support is used. When model errors dominate, it is thus better to sample the ``model space'' within some $\chi^2$ than the $\chi^2$-space with some fixed model.
As an example, shape models of the asteroid Daphne from adaptive optics images and photometry \citep{VK14} using both the octantoid representation and subdivision surfaces are shown in Fig.\ \ref{fig:daphne}. The models are quite similar and fit the data equally well, and their difference gives an idea of the real level of resolution. MCMC probing with either shape support leads to unrealistically small differences (insignificant compared to those in Fig.\ \ref{fig:daphne}).
Even the shape-support test is likely to produce too optimistic reliability limits; the model error can be further enlarged by, e.g., introducing random fluctuations in the scattering properties over the surface. This principle could be developed into a meta-level Monte Carlo procedure that probes the space of possible model types using latent parameters.

We conclude that shape sampling based on a fixed model type, no matter how diligently done with Monte Carlo or other methods, leads to overoptimistic resolution with artificial details. A typical example of this is the radar model of the asteroid Itokawa that portrayed imaginary detail at the resolution level expected from the data while not capturing even the large-scale features. There was nothing wrong with the model fit to the data as such: the inverse problem was simply nonunique (or very unstable) due to the restricted observing geometries and instrumental projection (Sect.\ 4.2), but the constrained shape support of the program did not reveal this \citep{OB05,No14}.

\subsection{Inversion with photometry only}
Since ADAM utilizes photometric data in addition to disk-resolved data, we note that ADAM can be used to reconstruct a model using photometric data only (simply by using only the photometric fit function from the toolbox). This is easy and fast to do (and the shape rendering is even faster than the standard convex inversion of lightcurves), but the result is inevitably unreliable: it is well known that even sizable nonconvex shape features require high solar phase angles to show in disk-integrated data \citep{KT01b,DK03,KD06}.  This can be probed with the shape reliability approach above. 

\begin{figure}
\begin{subfigure}[b]{0.15\textwidth}
                \includegraphics[clip=true,trim=200 310 180 250,scale=0.3]{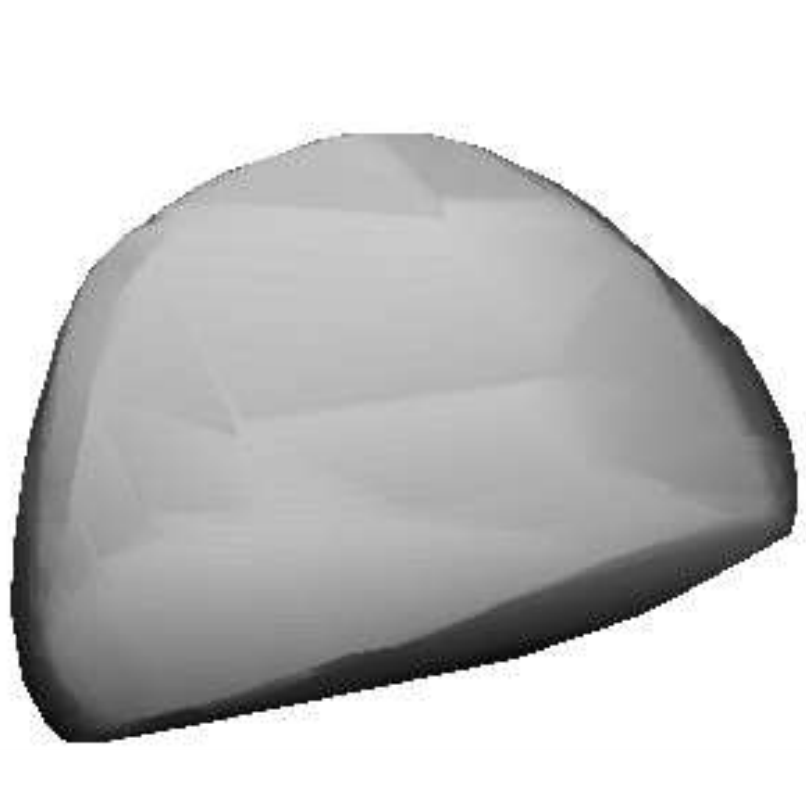}
        \end{subfigure}%
        \begin{subfigure}[b]{0.15\textwidth}
                \includegraphics[clip=true,trim=200 310 180 250,scale=0.3]{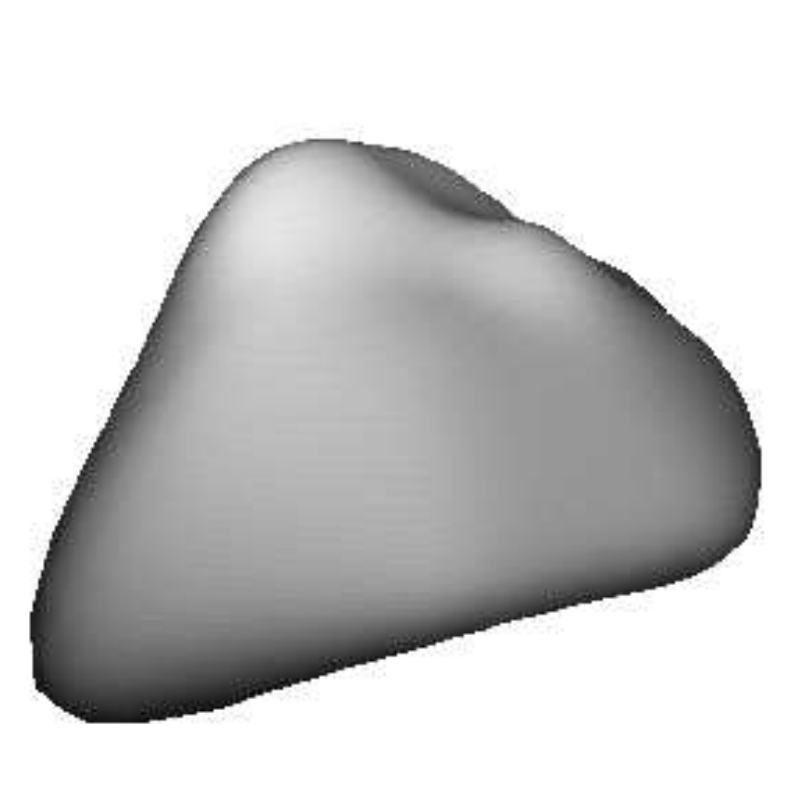}
        \end{subfigure}%
        \begin{subfigure}[b]{0.15\textwidth}
                \includegraphics[clip=true,trim=200 310 200 250,scale=0.3]{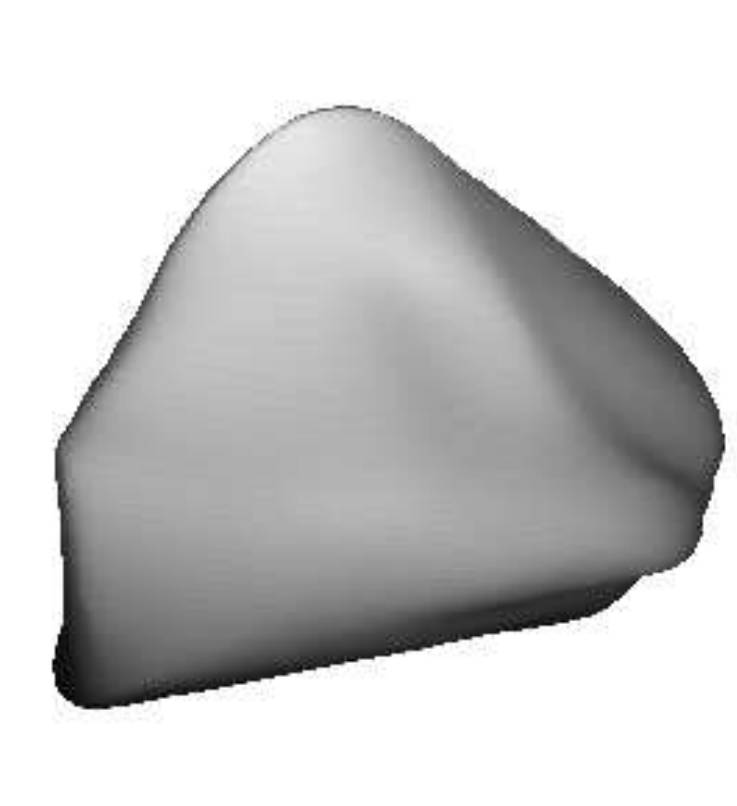}
        \end{subfigure}
\caption{\label{fig:golevka} Asteroid (6489)~Golevka reconstructed from disk-integrated photometry. From left to right: Convex, octantoid and subdivision surface.}
\end{figure}

As an example, we show reconstructed shapes of the asteroid Golevka in Fig.\ \ref{fig:golevka}, based on the data in \cite{KT01b}. Both the subdivision method and the octantoid-based model display additional detail not seen in the convex model. However, the details are not supported by the data: the convex model gives at least as good a fit as the nonconvex ones, as is almost always the case with lightcurves \citep[so far the only case of a better nonconvex model fit to photometry is that of the asteroid Eger in][]{DV12}. Indeed, with Golevka and other ground truth cases (maps from space probe missions), even the lightcurve fit with the correct shape and the scattering model assumed in inversion is not better than that with the convex model \citep{KT01b,KD06}. This underlines the fact that, due to systematic errors, any best-$\chi^2$ solution with photometry only is likely to miss the correct details. 

While the convex model yields the best global agreement with the radar-based Golevka model  \citep[see the comparison in Figs.~3 and 4 in][]{KM02}, the nonconvex ones portray much of the general sharpness and ruggedness of the body even though their details are not correct.  The convex shape presents something of a softened error envelope within which numerous local shape variations are possible (as if the target were seen unfocused), while the nonconvex representations are samplings of those variations. Their details coincide neither with each other nor with those in the radar-based model, but they are useful as illustrations and for probing the potential shape options \citep[cf. the nonconvex examples in][]{KP04}.

\section{Fourier transform and information content}

As discussed in \cite{VK14}, the Fourier transform (FT) facilitates a natural interpretation for the pixel size as the maximum frequency present in the data, and makes it easy to incorporate the impulse response function of the imaging system. It also makes the optimization procedure fast and straightforward, without the cumbersome aspects related to pixellated image fields and binned model image distributions. The principle of the ADAM approach is to compare, instead of the images themselves, a set of FT samples (typically some thousands depending on the level of resolution) from the model image with those of the data image, and to iterate until the best fit is  found. This is described in Sect.\ 5. 

Letting $\mathcal{T}$ be the set of facets forming a model polyhedron and $\mathcal{P}$ a projection operator, the two-dimensional Fourier transform of a projected polyhedron in the $(\xi,\eta)$-plane is
\begin{equation}
\label{FT1}\mathcal{F}(u,v)=\sum_{T_i\in\mathcal{T}}\iint_{\mathcal{P}T_i}B_i\,I_i(\xi,\eta)\,e^{-2\pi\imath(u\xi+v\eta)}\,\diff\xi\, \diff\eta,
\end{equation}
where $B_i$ is the luminosity value of the facet $i$, and the function $I(\xi,\eta)$ is unity if the point projected on $(\xi,\eta)$ is visible and zero otherwise. As shown in \cite{VK14}, we obtain by Green's theorem (dividing a facet into subfacets if necessary so that we may assume $I$ is constant within each subfacet)
\begin{equation}\label{foutrans}
\mathcal{F}(u,v)=\sum_i B_i \sum_j\, {\mathcal I}_{ij}(u,v),
\end{equation}
where 
\begin{equation}
\begin{split}
{\mathcal I}_{ij}(u,v)=\frac{1}{4\pi^2(u^2+v^2)}\frac{(b-d)u-(a-c)v}{(a-c)u+(b-d)v}\\
\times\,\left[e^{-2\pi\imath (au+bv)}-e^{-2\pi\imath (cu+dv)}\right]
\end{split}
\end{equation}
for the $j$-th boundary line segment (oriented counterclockwise) of the facet $i$, with the end points $(a,b)$ and $(c,d)$.

The summation over the interior edges of a projected polyhedron can be reordered by noting that each polygon edge in the interior is shared by two polygons, so a new factor $\tilde B$ can be taken to be the difference between the two $B_i$, and the edge term is computed only once. 
This explicitly shows why most of the information in the image is indeed from the limb and shadow boundary curves discussed in \cite{Ka11} and \cite{KV12}. The values of $\tilde B$ for interior triangle edges are usually close to zero (indeed, they vanish for the geometric scattering $B_i=const.$), so most of the weight is on the boundary edges. In practice, this is confirmed by the similar results for, e.g., the asteroid Daphne obtained by KOALA and ADAM. There is little real information in the interior pixels of adaptive optics images, but on the other hand their errors do not distort the result either: the difference between the KOALA and ADAM models (for the same initial values and shape support) is negligible.

The role of boundary information can be understood when comparing to the extreme case of lightcurve data: if we sum the pixel brightnesses over the image as in photometry, all the local shape information in the image is lost, so the remaining information is considerably more dependent on the light-scattering properties that are never very well known. But with images the boundary contrast is always the largest, so it is sufficient to have some kind of reasonable scattering (or thermal distribution) model to account for the interior pixel contrasts. Indeed, the uniqueness theorems on the image, interferometry, occultation, or radar data are based on the robust boundary contour information \citep{Ka11,KV12,VK14}. With disk-integrated data only, Minkowski stability is luckily on our side when using convex models \citep{KT01b,KM02}.

\section{Data sources}
The versatility of the ADAM algorithm allows the handling of different data sources with only minor changes to the instrument-dependent part of the procedure (essentially just the definition of the instrumental projection plane and the adopted point-spread function). In this section, we present diverse examples of shape reconstruction with ADAM using both simulated and observed data.

\subsection{Interferometry and ALMA}
The interferometric imaging method differs radically from a typical telescope; instead of observing the sky brightness directly, the interferometer samples the Fourier transform of sky brightness. Each antenna pair of the interferometric array determines one sample on the Fourier plane. The maximum separation between antennas determines the maximum attainable resolution. 
The interferometer most relevant to asteroid shape studies is the Atacama Large Millimeter Array (ALMA) in the Chilean desert. In its full configuration, the interferometer will be capable of observing at the resolution of a few milliarcseconds at the wavelength of 0.3\,mm, corresponding to the separation of 16\,km between antennas.

\begin{figure}
 \centering   
        \begin{subfigure}[b]{0.25\textwidth}        
                \includegraphics[clip=true,trim=50 10 50 20,scale=0.3]{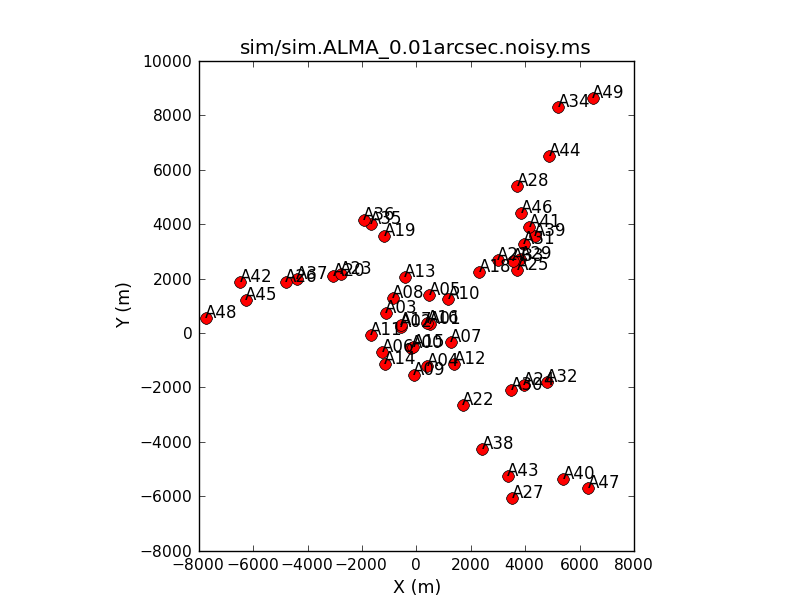}
        \end{subfigure}%
        \begin{subfigure}[b]{0.25\textwidth}           
                \includegraphics[clip=true,trim=50 10 50 20,scale=0.3]{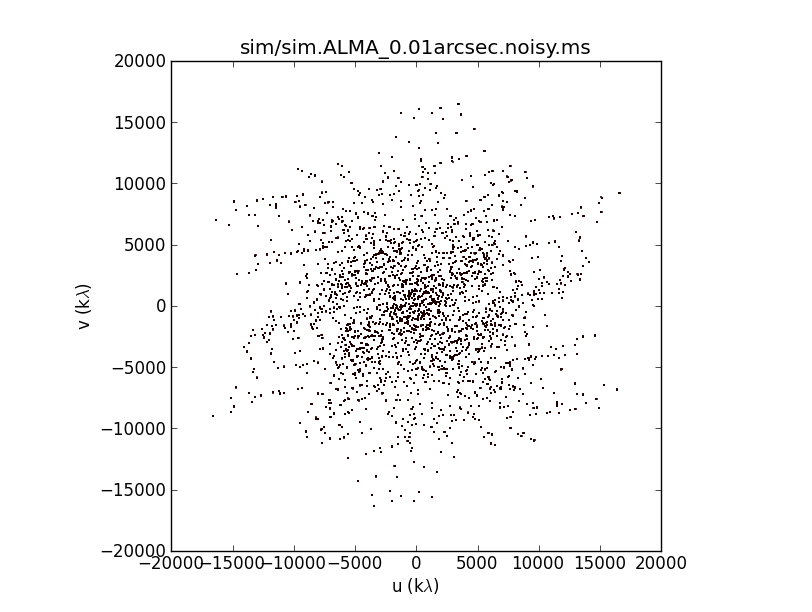}
        \end{subfigure}
        \caption{ \label{fig:inter_antenna}The antenna locations of ALMA (left) and corresponding uv-plane visibilities (right). Images generated with the CASA software package.}
        \end{figure}

Given the brightness distribution $I(\xi,\eta)$ on the plane-of-sky, the visibility function is defined as the integral
\begin{equation}
V(u,v)=\iint I(\xi,\eta)\,e^{-2\pi\imath(u\xi+v\eta)}\,\diff\xi\, \diff\eta,
\end{equation}
which is a two-dimensional Fourier transform of the brightness distribution. Each antenna pair, corresponding to the projected baseline on the plane-of-sky, samples the visibility function. When the visibility function is sampled on a sufficiently dense set, the Fourier transform can be inverted to obtain the brightness distribution $I(\xi,\eta)$. But since the function $V(u,v)$ is measured only at a finite number of points, the observed visibility function is 
\begin{equation}
\tilde{V}(u,v)=F(u,v)\,V(u,v),
\end{equation}
where $F(u,v)$ is a sampling function corresponding to the sampled points on the $(u,v)$-plane. Thus the obtained brightness distribution is actually
\begin{equation}
\tilde{I}(\xi,\eta)=f(\xi,\eta)\star I(\xi,\eta),
\end{equation}
i.e., a convolution of the true brightness distribution with the inverse Fourier transform $f(\xi, \eta)$ of the sampling function. Deducing the true brightness distribution $I$ from the partially measured brightness $\tilde{I}$ is an inverse problem and there are several iterative algorithms to infer $I$ from $\tilde{I}$, see, e.g., \cite{LL06}.

While the images obtained from the interferometer are informative, the great advantage with ADAM is that the algorithm works directly with the values of the visibility function obtained from the instrument. This approach has several distinct advantages:
\begin{itemize}
\item Sparse data may be used (e.g. interferometry with a few baselines)
\item The distribution of antennas does not cause bias, since the Fourier transform is not inverted
\item Possible artefacts caused by the inversion process are avoided
\item The dependence between different observations is taken automatically into account 
\end{itemize}   

\begin{figure}
\begin{subfigure}[b]{0.3\columnwidth}
\includegraphics[clip=true,trim=90 40 90 30,scale=0.3]{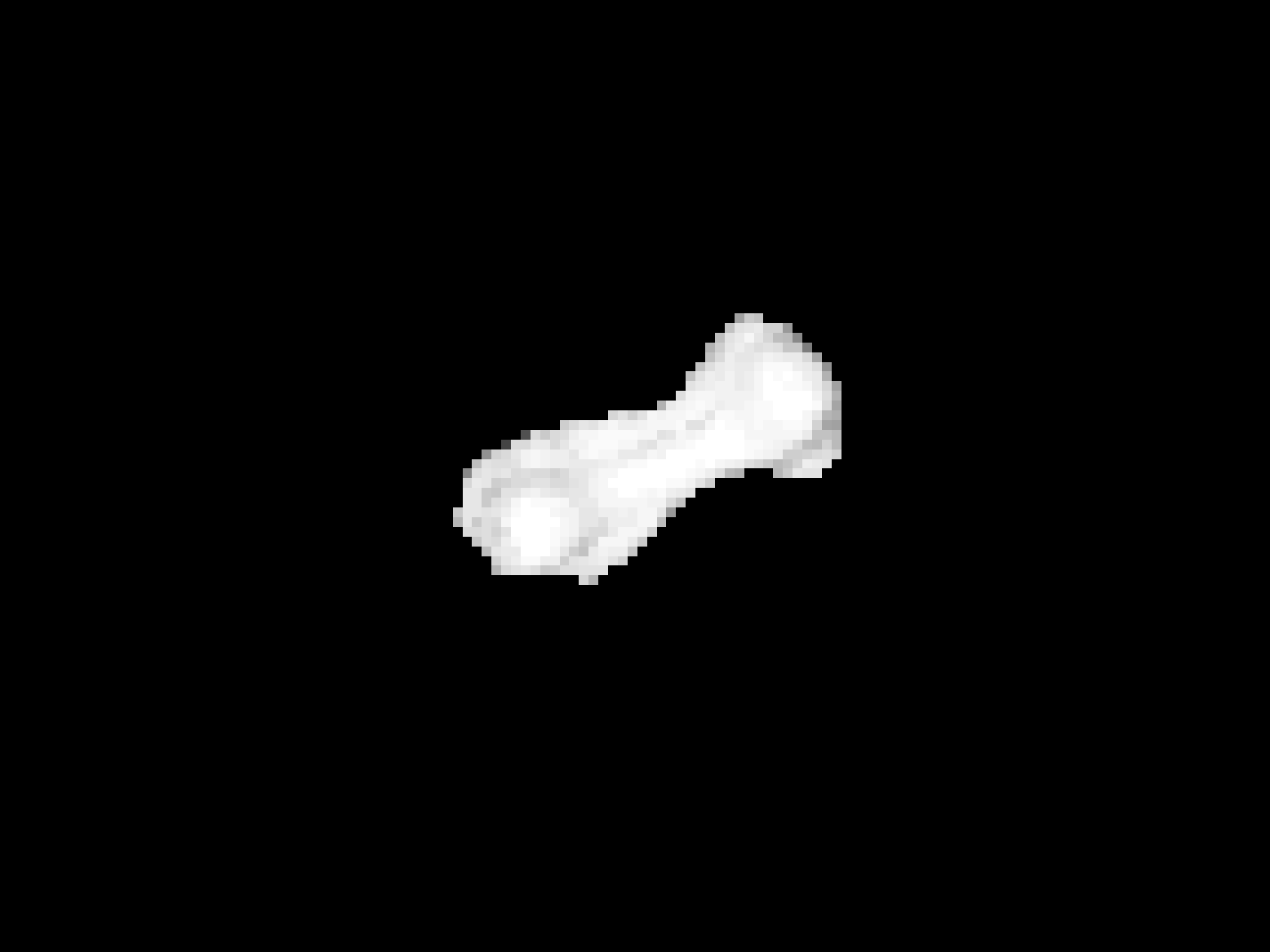}
\end{subfigure}
\begin{subfigure}[b]{0.3\columnwidth}
\includegraphics[clip=true,trim=90 40 90 30,scale=0.3]{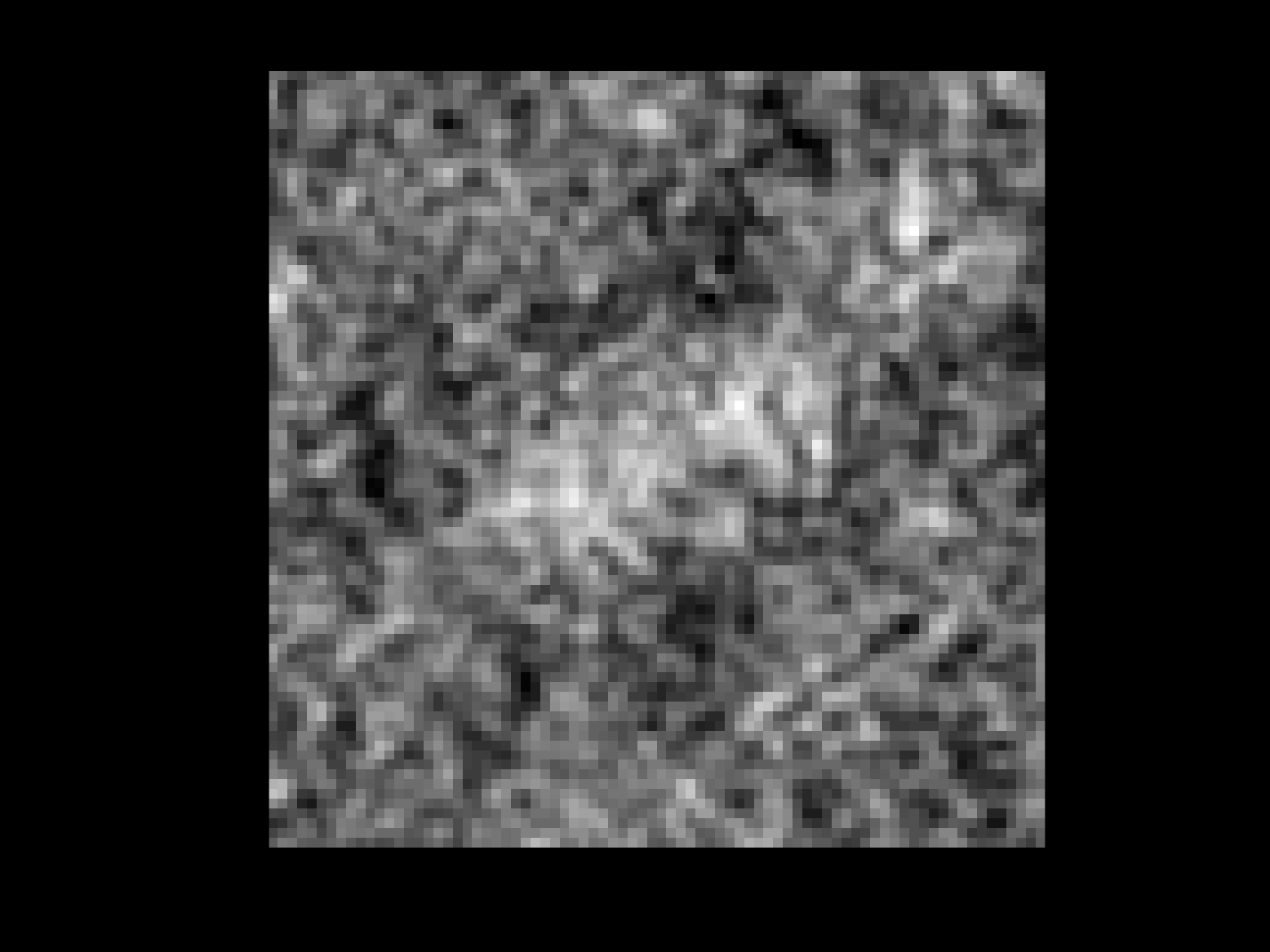}
\end{subfigure}
\begin{subfigure}[b]{0.3\columnwidth}
\includegraphics[clip=true,trim=90 40 90 30,scale=0.3]{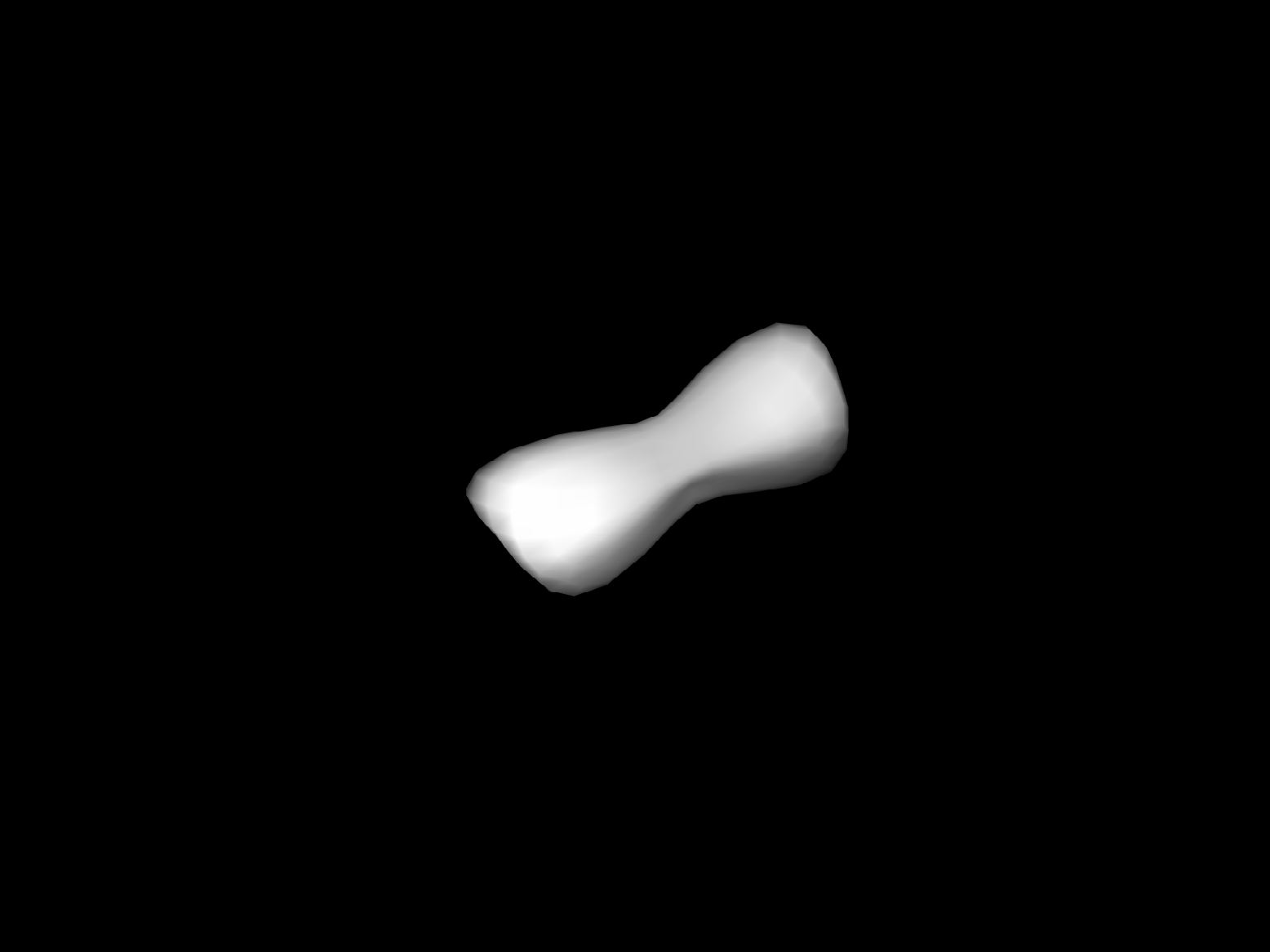}
\end{subfigure}

\begin{subfigure}[b]{0.3\columnwidth}
\includegraphics[clip=true,trim=90 40 90 30,scale=0.3]{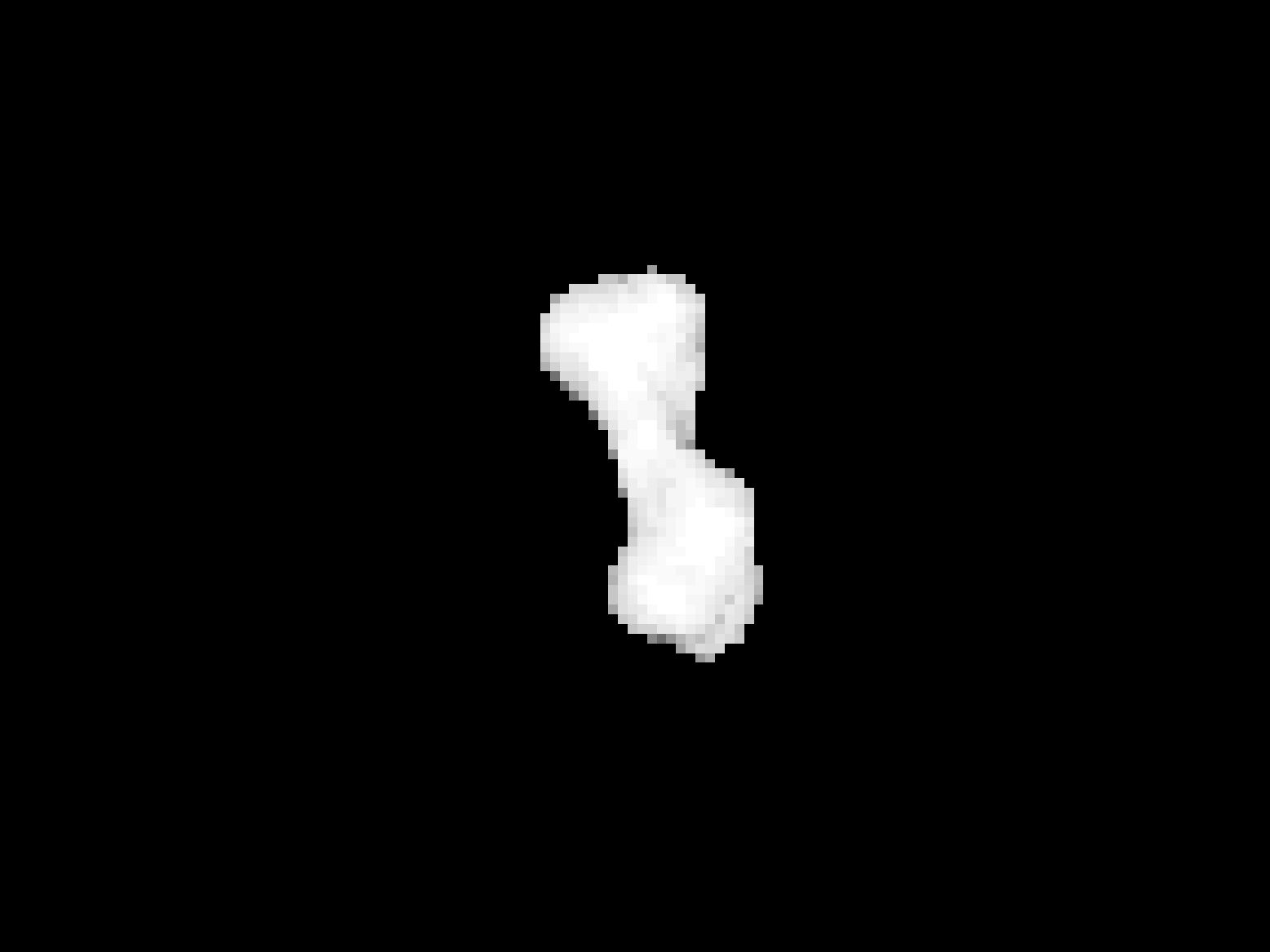}
\end{subfigure}
\begin{subfigure}[b]{0.3\columnwidth}
\includegraphics[clip=true,trim=90 40 90 30,scale=0.3]{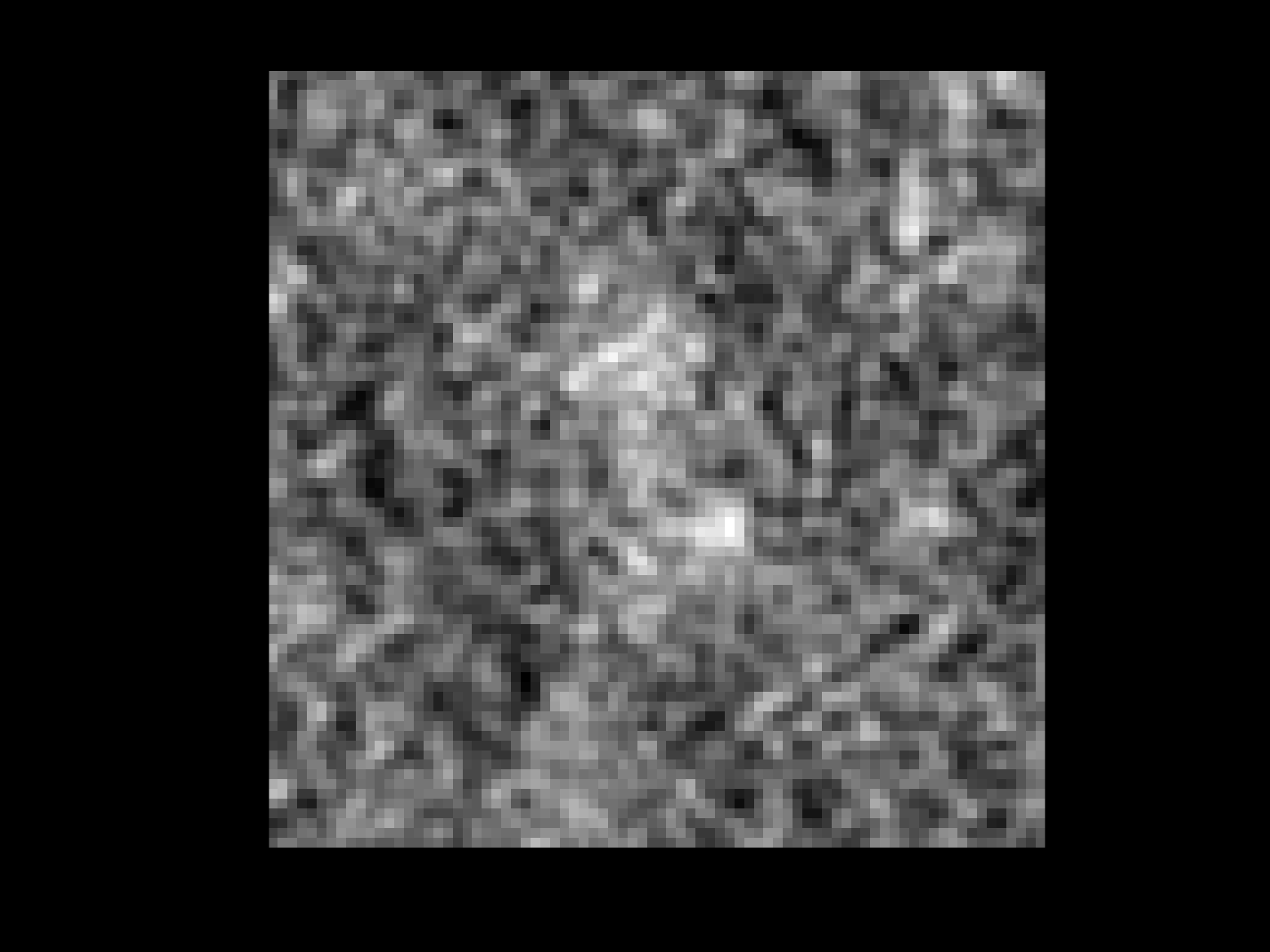}
\end{subfigure}
\begin{subfigure}[b]{0.3\columnwidth}
\includegraphics[clip=true,trim=90 40 90 30,scale=0.3]{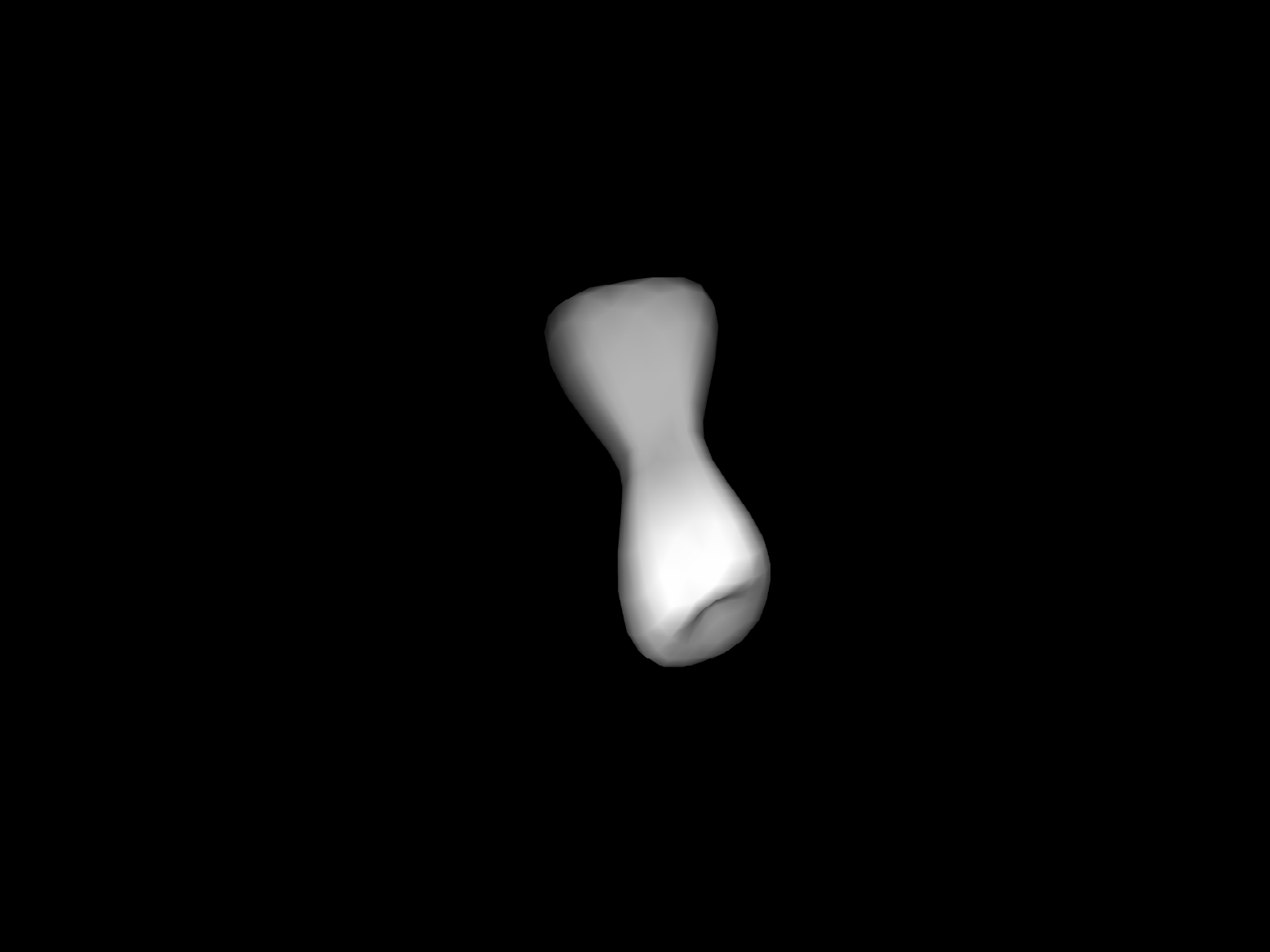}
\end{subfigure}

\begin{subfigure}[b]{0.3\columnwidth}
\includegraphics[clip=true,trim=90 40 90 30,scale=0.3]{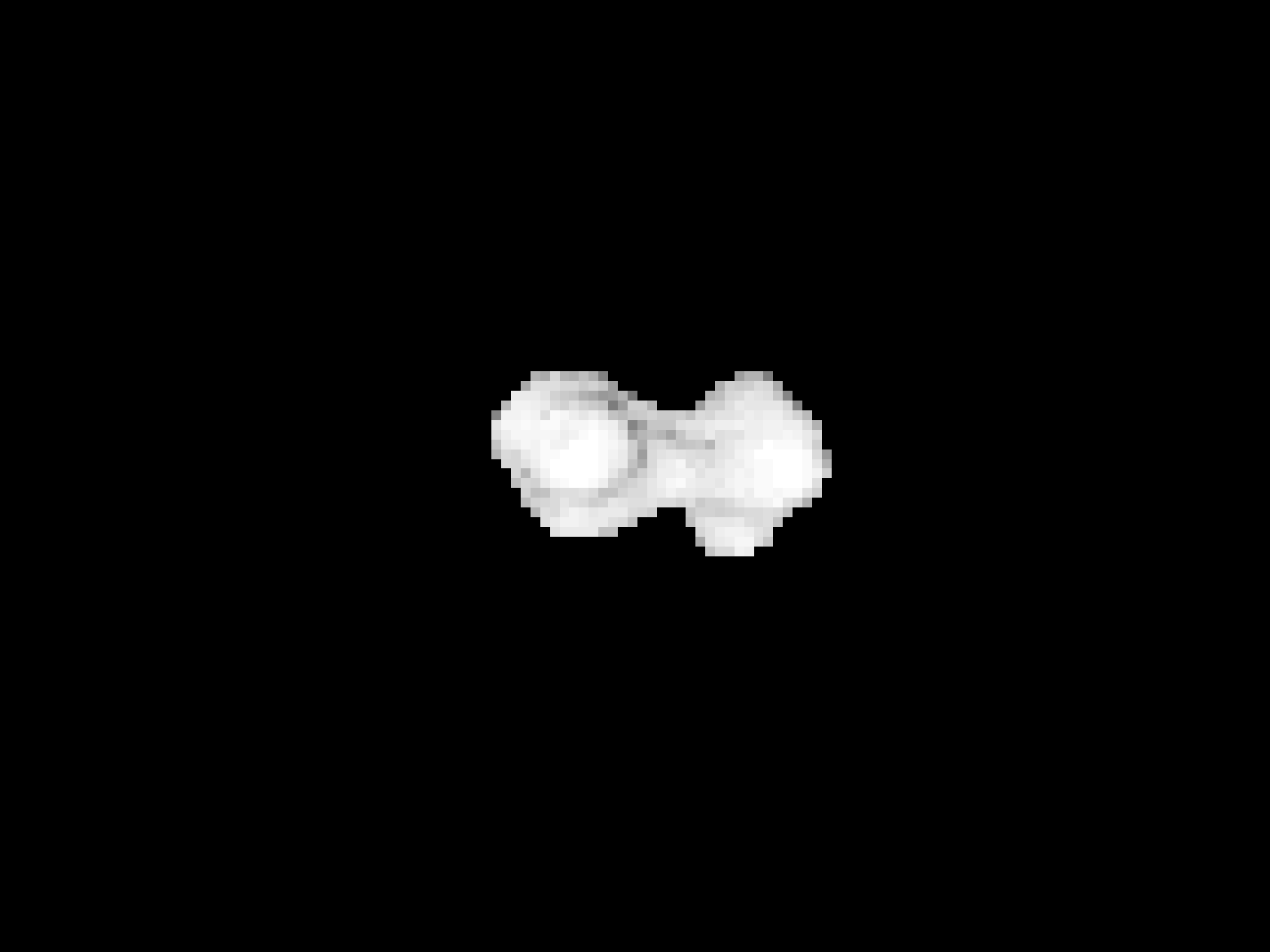}
\end{subfigure}
\begin{subfigure}[b]{0.3\columnwidth}
\includegraphics[clip=true,trim=90 40 90 30,scale=0.3]{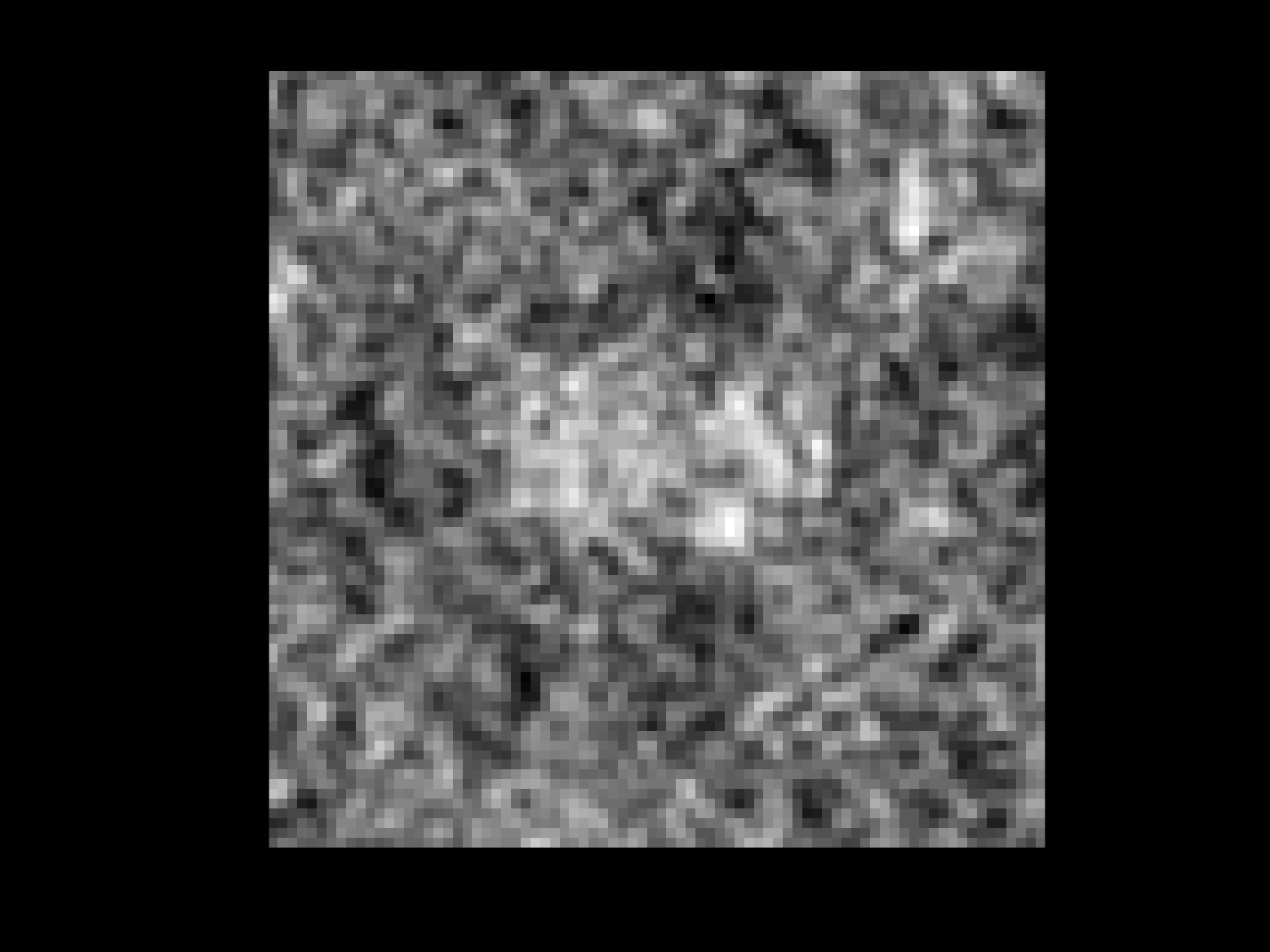}
\end{subfigure}
\begin{subfigure}[b]{0.3\columnwidth}
\includegraphics[clip=true,trim=90 40 90 30,scale=0.3]{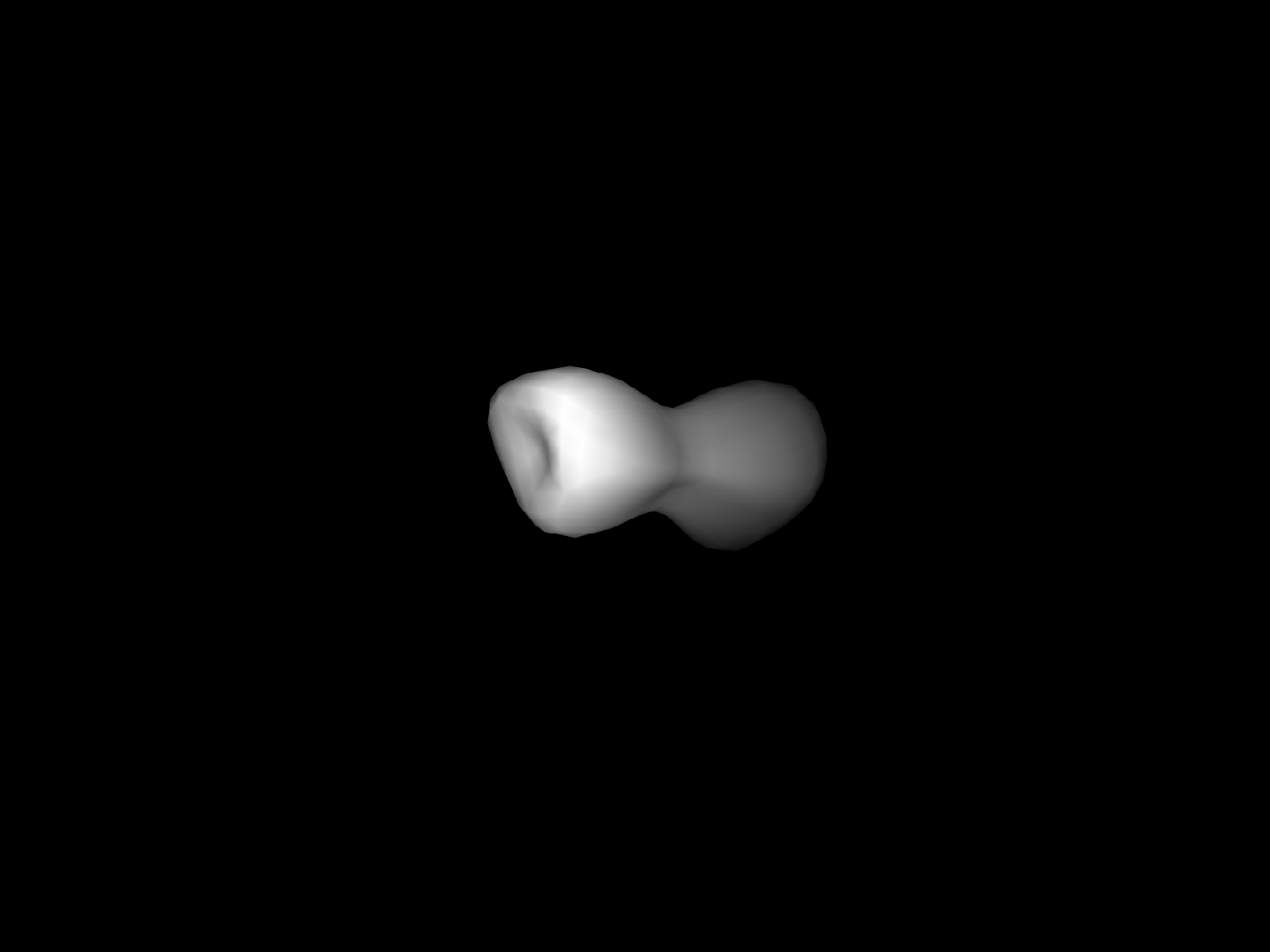}
\end{subfigure}



\begin{subfigure}[b]{0.3\columnwidth}
\includegraphics[clip=true,trim=90 40 90 30,scale=0.3]{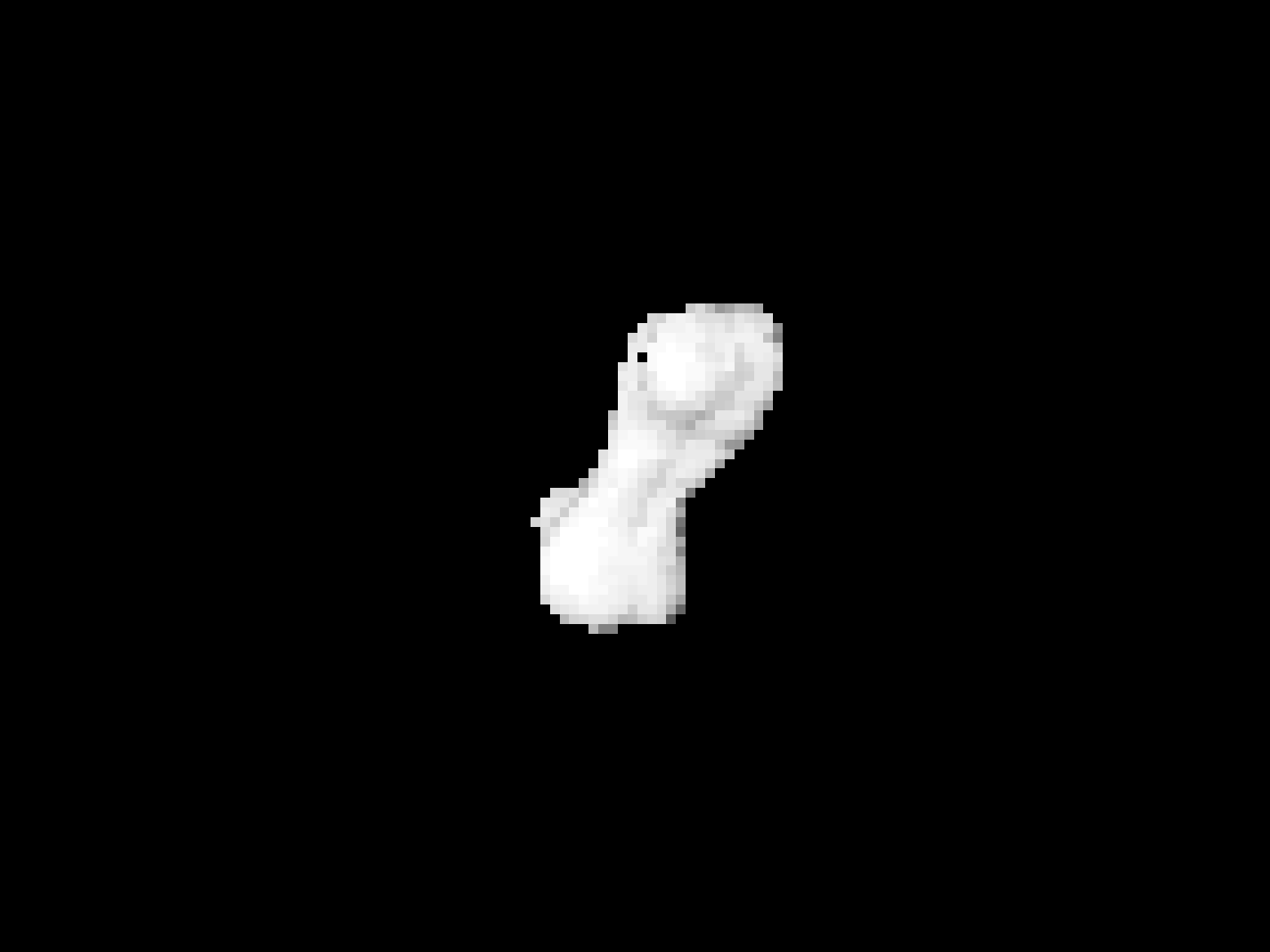}
\end{subfigure}
\begin{subfigure}[b]{0.3\columnwidth}
\includegraphics[clip=true,trim=90 40 90 30,scale=0.3]{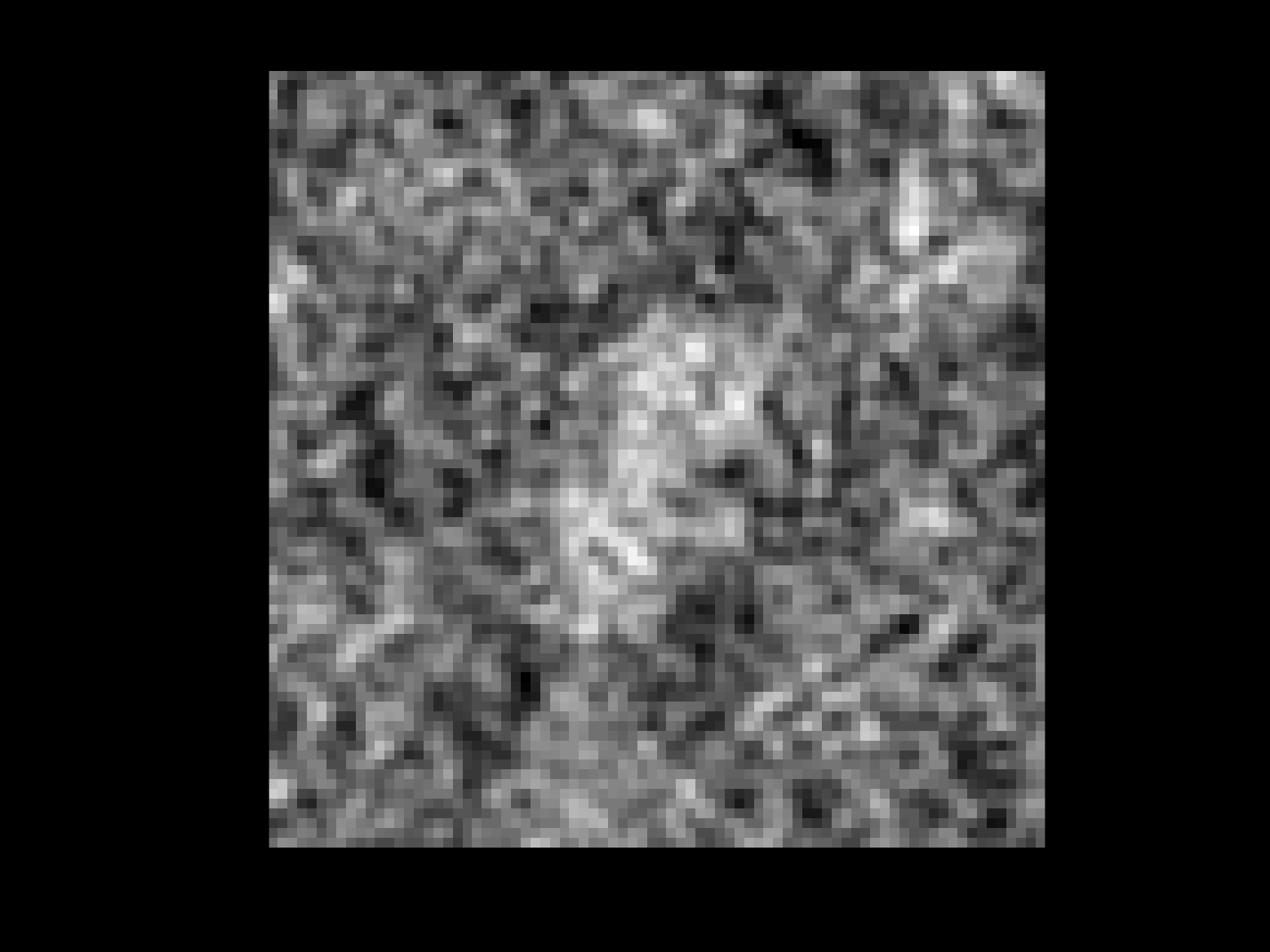}
\end{subfigure}
\begin{subfigure}[b]{0.3\columnwidth}
\includegraphics[clip=true,trim=90 40 90 30,scale=0.3]{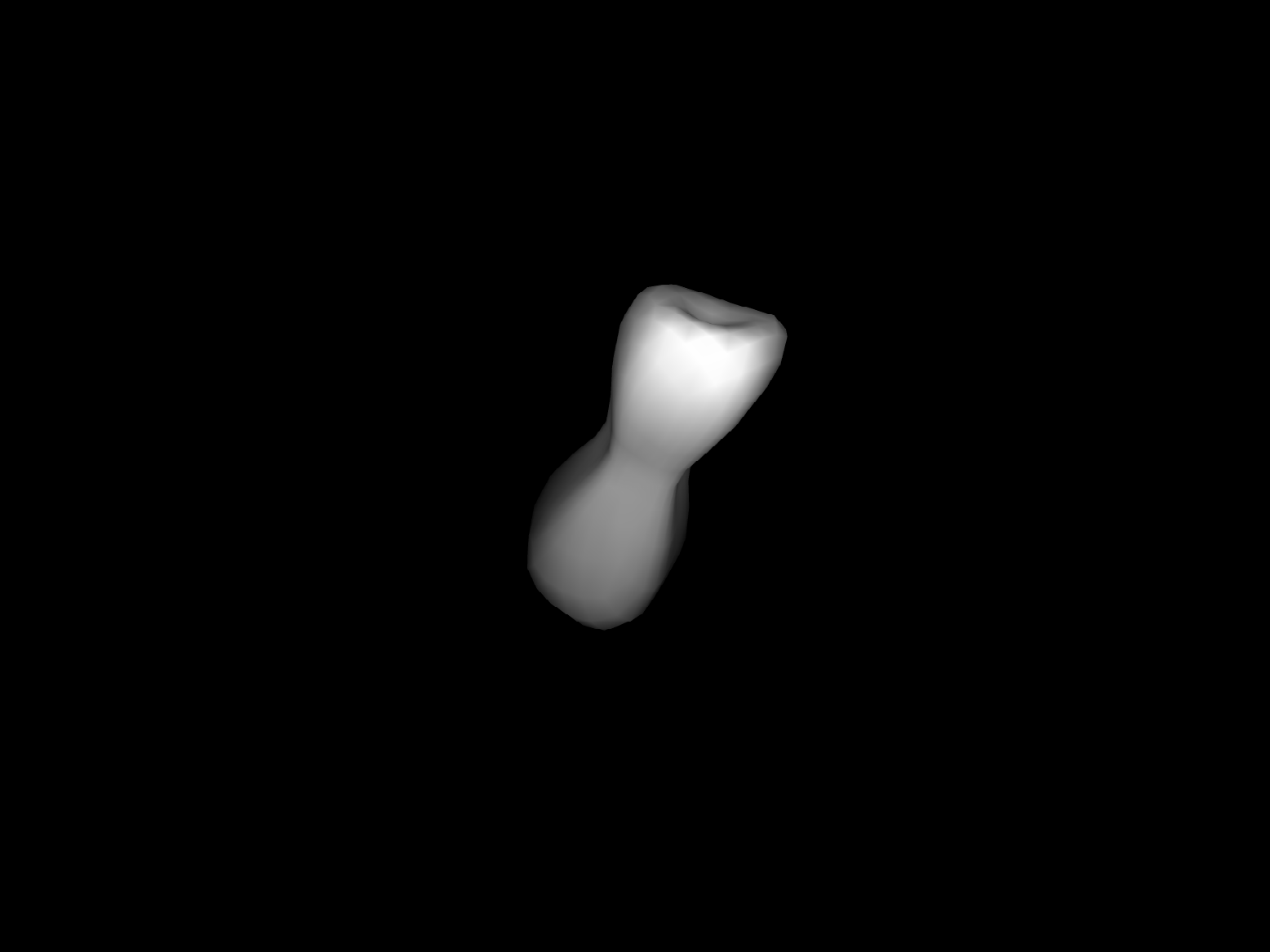}
\end{subfigure}

\begin{subfigure}[b]{0.3\columnwidth}
\includegraphics[clip=true,trim=90 40 90 30,scale=0.3]{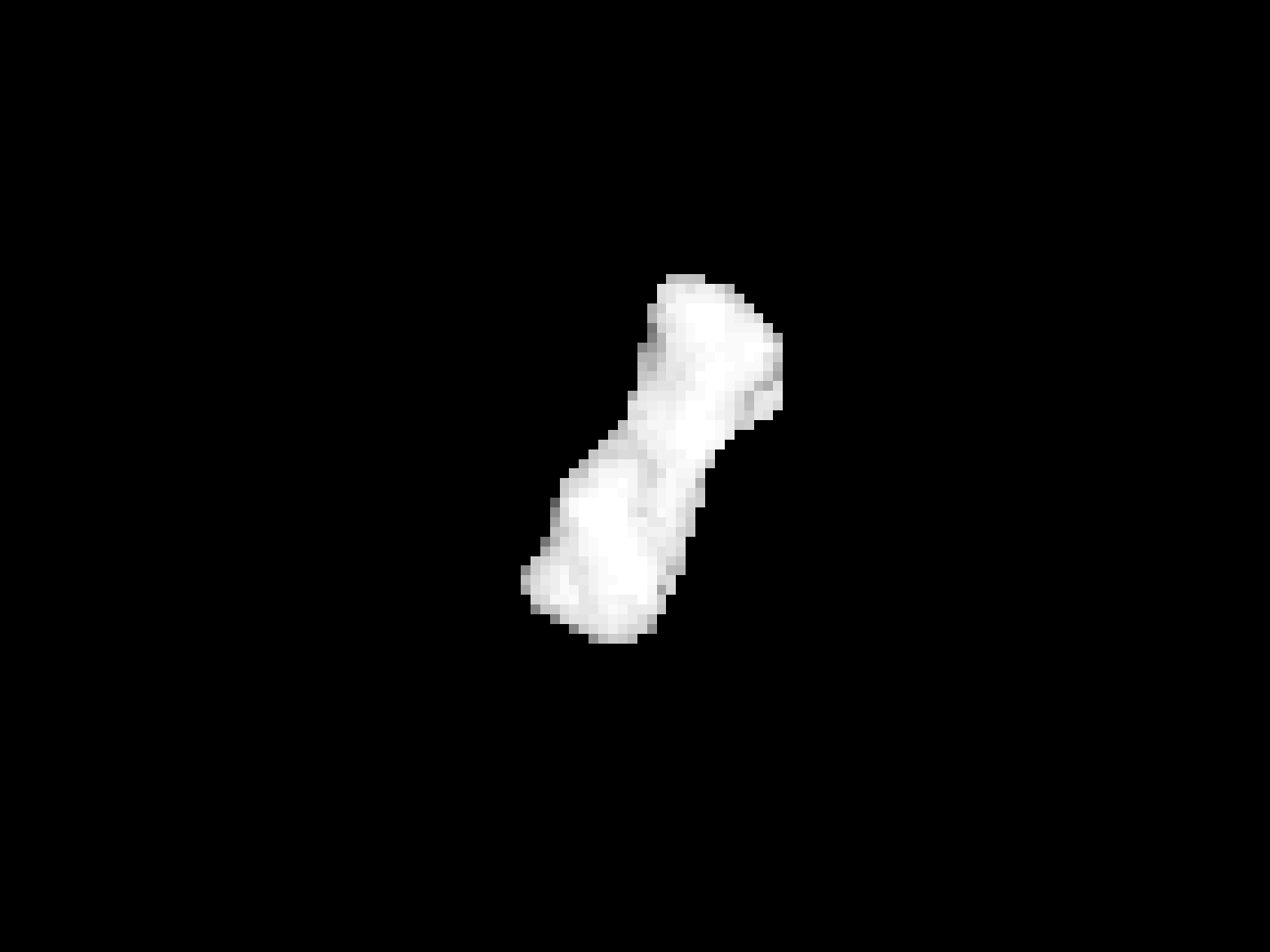}
\end{subfigure}
\begin{subfigure}[b]{0.3\columnwidth}
\includegraphics[clip=true,trim=90 40 90 30,scale=0.3]{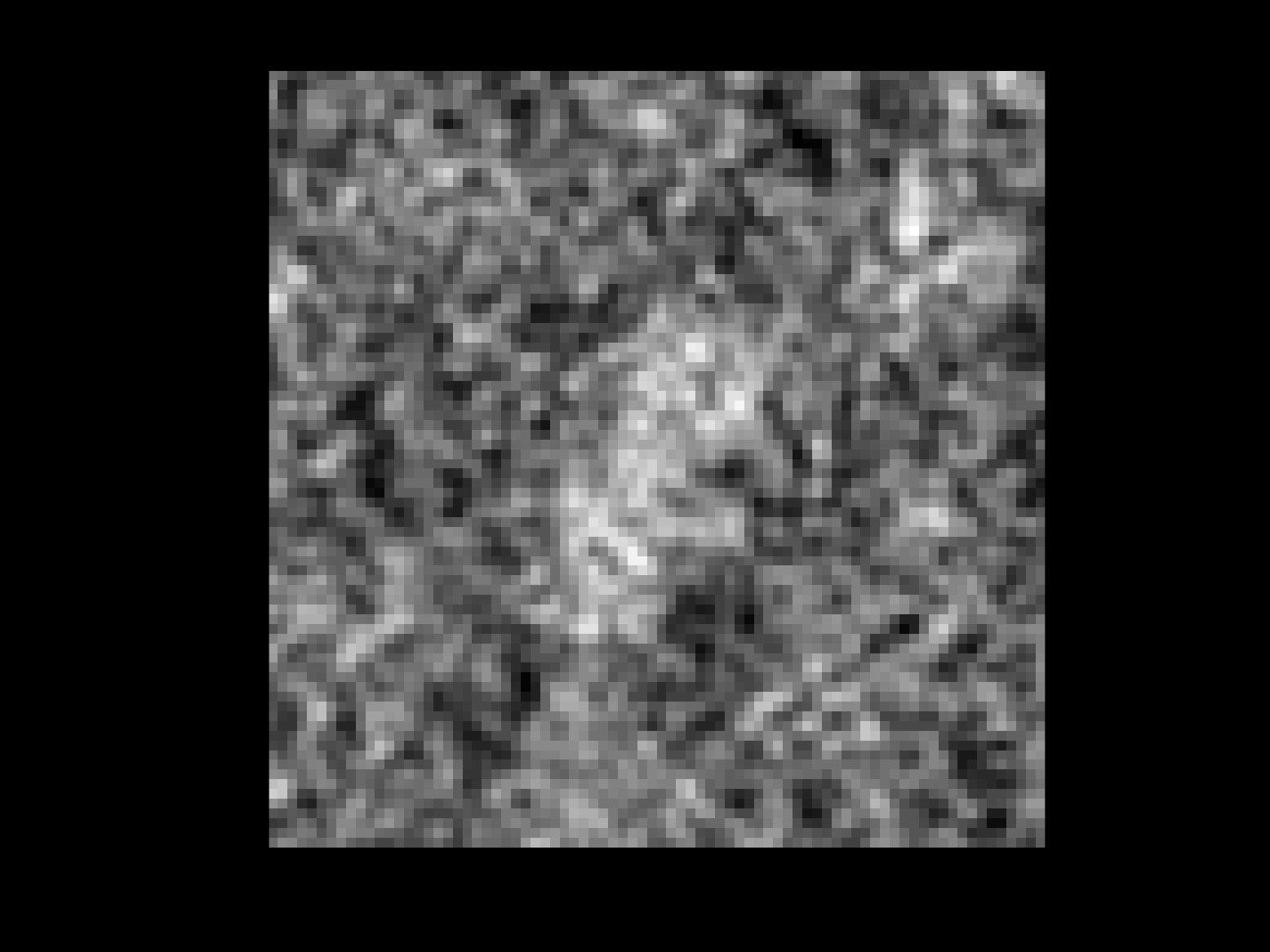}
\end{subfigure}
\begin{subfigure}[b]{0.3\columnwidth}
\includegraphics[clip=true,trim=90 40 90 30,scale=0.3]{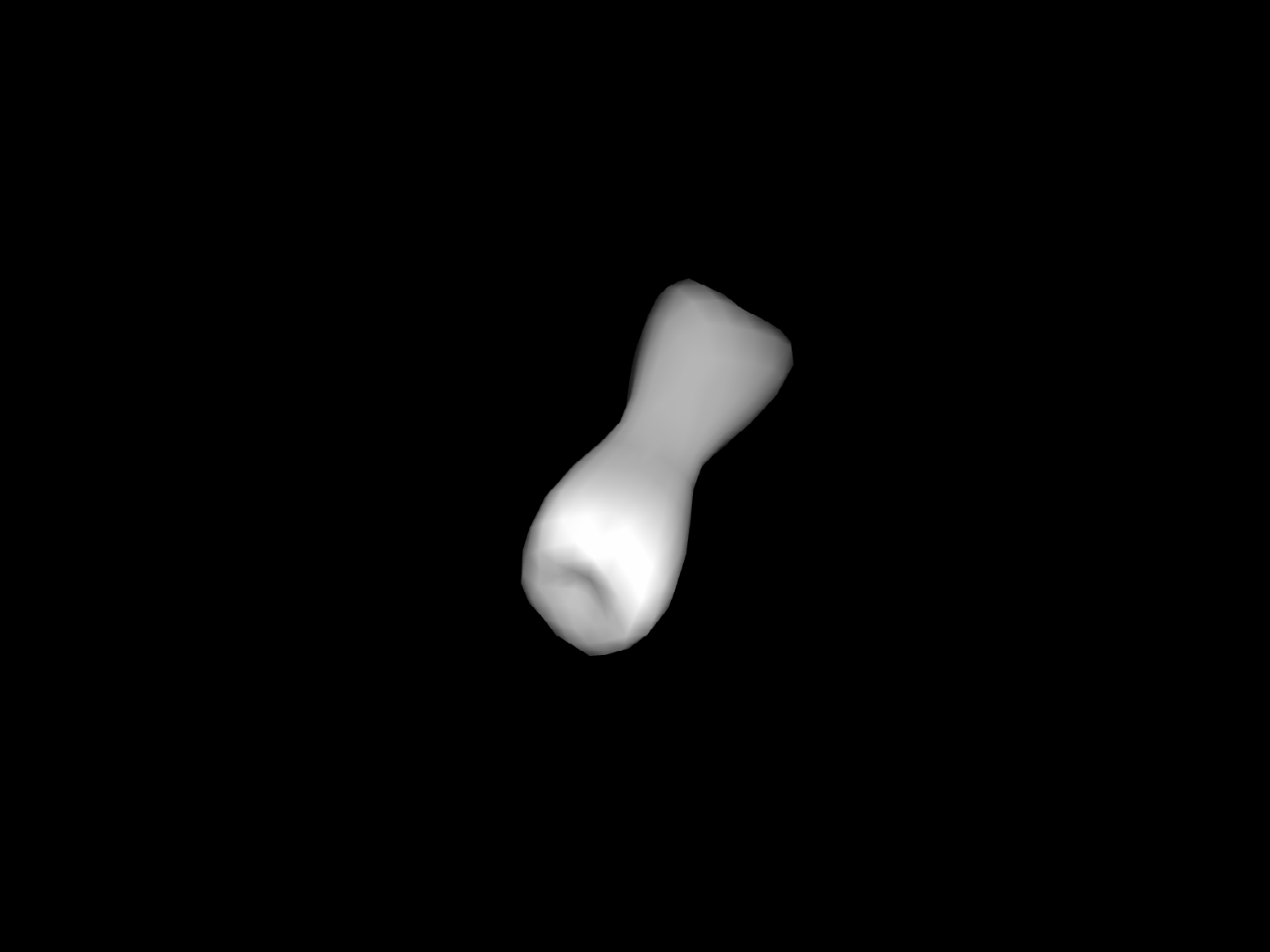}
\end{subfigure}
\caption{\label{fig:inter_pics} Simulated, uncorrupted images with 5 mas pixel size (left column), observed dirty images generated with CASA (middle) and the reconstructed low-resolution shape model (right). Note that the middle-column images are not needed in inversion; we use the direct FT data instead. The images are what would be seen if the raw data were deconvolved for viewing purposes as is usually done for ALMA targets. The test shape model is from \protect\cite{Ostro00}.}  
\end{figure}

To obtain the luminosity values for the model surface (i.e., the brightness factor $B_i$ for each facet) in the infrared regime of ALMA, we can use the Fourier-series approximation of \cite{NV08} as discussed in \cite{VK14}. The fast analytical computations are then efficient in the optimization.  
A  simple thermophysical model is sufficient for shape reconstruction, as the most relevant information is contained in the boundary data, which are quite robust with respect to the thermal model used.  This is in  contrast to the disk-integrated thermal data that are more sensitive to both the surface properties and the thermal model. 

For thermal infrared imaging, ALMA facilitates asteroid observations at resolution levels previously attained only by range-Doppler radar. To explore the possibilities of ALMA for shape modelling, we use the Common Astronomy Software Applications (CASA) package developed by National Radio Astronomical Observatory (NRAO) to simulate observations.

Consider a hypothetical asteroid with geocentric and heliocentric distances of 1.5 and 1 AU, respectively. The thermal flux is observed at the 350 GHz band, a frequency located in an atmospheric window. There are 11 observation runs, each observation lasting 50 s with ten-second integration time.  We choose an antenna configuration providing approximate resolution of 10 mas, a resolution which is well within the capabilities of ALMA. The antenna configuration and the corresponding uv-plane sampling pattern are shown in Fig. \ref{fig:inter_antenna}.  Uncorrupted plane-of-sky images, with a resolution of five milliarcseconds, are displayed in the column on the left in Fig. \ref{fig:inter_pics}. We use the CASA software to add realistic atmospheric noise to the observations. The resulting dirty images, which are obtained by assuming that the unsampled frequencies are zero, are shown in the middle column. These images are provided for illustration purposes only, since ADAM uses the $uv$-plane samples directly.

To test the ADAM reconstruction method, we use a low-resolution octantoid representation with 75 shape parameters. We also fit a scaling term, common to all observations. Usually it is a good idea to use scaling specific to each observation, but in this case we know that all the simulated observations are done in similar conditions, so the common scaling term is justified. The reconstructed shape is displayed in the right column in Fig.\ \ref{fig:inter_pics}, with the same observation geometries as for the model images. The small-scale detail is lost, which is to be expected due to the added atmospheric noise and coarse instrument resolution. However, the bifurcated shape is well recovered despite the noisy data (note that we used ALMA data only). The computation time for this reconstruction was a few minutes. For real observations, complementary data are often provided by other observation methods, e.g., disk-integrated photometric data are almost always available.

\subsection{Radar data}
The mathematical principles of the feasibility and uniqueness of the inversion of range-Doppler images are discussed in \cite{VK14}. Here we consider some practical issues related to shape reconstruction.
While other imaging methods rely on detecting the radiation of the sun that is  reflected or re-radiated from the asteroid, radar provides its own illumination, making it possible to observe an asteroid regardless of the position of the sun. Moreover, in contrast to the visible or infrared wavelengths, the frequencies used by the radar are not significantly distorted by the atmosphere. Additionally, the properties of the waveform may be carefully controlled to reveal structural details on the surface of the asteroid. These properties make it possible to obtain \emph{data resolution} down to ten meters or less for near-Earth asteroids, but this does not immediately translate to the same \emph{model resolution} because of the inverse problem (cf.\ the Itokawa example in Sect.\ 2.3). 

Range-Doppler radar resolves an object both in the range and in the line-of-sight velocity that translates to the Doppler shift of the reflected pulse. The frequency spectrum may be extracted by taking the fast Fourier transform of the pulses corresponding to a particular range gate. The actual hardware implementation and the signal processing are complicated as the detected signals are below the noise level of the instrument \citep{OH02}. Fortunately, the technical specifics are not required for the actual shape reconstruction, since the radar performance may be modelled by the point-spread function of the system.

The point $p=(x,y,z)$ on the asteroid's surface can be transferred  to the range-Doppler frame $(r,D)$ by the linear mapping 
\begin{eqnarray}\label{radarframe} 
r&=&(x \cos\phi+y\sin\phi)\sin\theta+z\cos\phi\\
D&=&\omega\sin\theta\,(x\sin\phi-y\cos\phi),
\end{eqnarray}
where $\omega$ is the rotation rate of the asteroid around the $z$-axis and $(\theta,\phi)$ are the spherical radar direction coordinates as seen from the asteroid.  In this mapping, the range-Doppler radar image brightness $L$ may be written as an integral over the asteroid surface $S$:
\begin{equation}\label{radarim} 
L(r,D)=\iint_S h_r\left[r-r(p)\right]\, h_D\left[D-D(p)\right]\, B(p)\, I(p) \,\diff S  ,
\end{equation}
where $h_D$ and $h_r$ are the point-spread functions of the radar system, corresponding to the Doppler-shifted frequency $D$ and the range $r$, respectively. Here $I$ is the visibility function, which is unity if the point is visible to the radar and zero otherwise. This form is similarly defined for all generalized projections \citep{KL06}. The mapping $p\rightarrow (r,D)$ is unique, but its inverse is many-to-one, so the inherent information content of a range-Doppler image is considerably smaller than that of an optical image of similar resolution. Thus, while the nominal resolution provided by radar may be unmatched by any other instrument, the drawback of radar imaging is the difficulty of the interpretation of the images.

The radar scattering function is given by $B$, which is usually a simple cosine law
\begin{equation}\label{eqcoslaw} 
B(p)=C\left[\mu(p)\right]^n,
\end{equation}
where $\mu$ is the cosine of the angle between the surface normal and the radar direction. The constants $C$ and $n$ measure the surface reflectivity and the specularity of scattering, respectively. The validity of Eq.\ (\ref{eqcoslaw}) for modelling the microwave scattering from the asteroid's surface is a rather convoluted question. While the cosine law is quite simplified, it should be noted that as the reflected wave is formed in a complicated manner by the surface material whose properties and roughness are usually unknown, fully realistic modelling of the reflected wave is not computationally feasible. However, as in the other disk-resolved cases, most of the information is contained in the boundary contours and is thus independent of the scattering model used.
 
Assuming the asteroid is modelled as a polyhedron with triangular facets $\mathcal{T}$, the integral (\ref{radarim}) may calculated separately for each facet, after projecting each triangle $T_i$ as a triangle $\mathcal{P}T_i$ on the range-Doppler plane:
\begin{equation}
L(r,D)=\sum_{T_i\in \mathcal{T}}  B_i I_i\int_{\mathcal{P}T_i} h_r(r-r')\, h_D(D-D')\, \diff r'\, \diff D',
\end{equation}
where we have assumed that the visibility $I$ and the scattering law $B$ are constant within a triangle.

Taking the Fourier transform on both sides, applying the convolution theorem, and writing $\mathcal{T}_i(u,v)$ for the sum over the edges of a Fourier transformed triangle as in Sect.\ 3, we obtain 
\begin{equation}
\mathcal{L}(u,v)=\sum_{T_i\in\mathcal{T}} B_i\, I_i\, H_r(u)\,H_D(v)\,\mathcal{T}_i(u,v),
\end{equation}
where $H_r(u)$ and $H_D(v)$ are, respectively, the Fourier transforms of $h_r$  and $h_D$. 

Like any images, radar plots are seldom correctly aligned in some reference frame due to the errors in the center of mass prediction, so the actual position of a radar image with respect to the two-dimensional projection of the model must be determined during the optimization. The task of image alignment is further complicated by the peculiar asymmetric structure of radar images, especially the bright leading edge, other possible ridges of strong reflectivity, and the fading farthest-range pixels.   If the alignment information is unknown, it is usually a good idea to fit the image offsets first to a fixed shape, obtaining better initial positions that can be used in the shape optimization.      

To demonstrate the reconstruction method, we make a fast ADAM model of the near-Earth asteroid 2000 ET$_{70}$. Our goal is to get a quick first look at an initial model (to be refined at will). The asteroid was observed during February 2012  at Arecibo and Goldstone observatories using 2380 and 8560\,MHz range-Doppler radars \citep{Na13}. The images obtained from Arecibo have a resolution of 15\,m in  range and 0.075\,Hz in frequency. Goldstone images have a somewhat lower resolution, 15 to 75 m and 1\,Hz, respectively. Our goal is to produce medium-scale detail in the reconstructed shape, so a typical model choice is an octantoid with $l_{\rm max}\sim 10$ and around 1500 facets. Our example is ``first-result oriented'' on purpose, so we assume no information about the instrument-specific distortions, or more importantly, knowledge about the point-spread functions determined by the instrument and the processing routines of the radar signal. Thus the point-spread function used in the shape reconstruction is simply the two-dimensional delta function. 

\begin{figure}
 \centering
  \begin{subfigure}[b]{0.3\columnwidth}
                \includegraphics[clip=true,trim=140 300 100 250,scale=0.25]{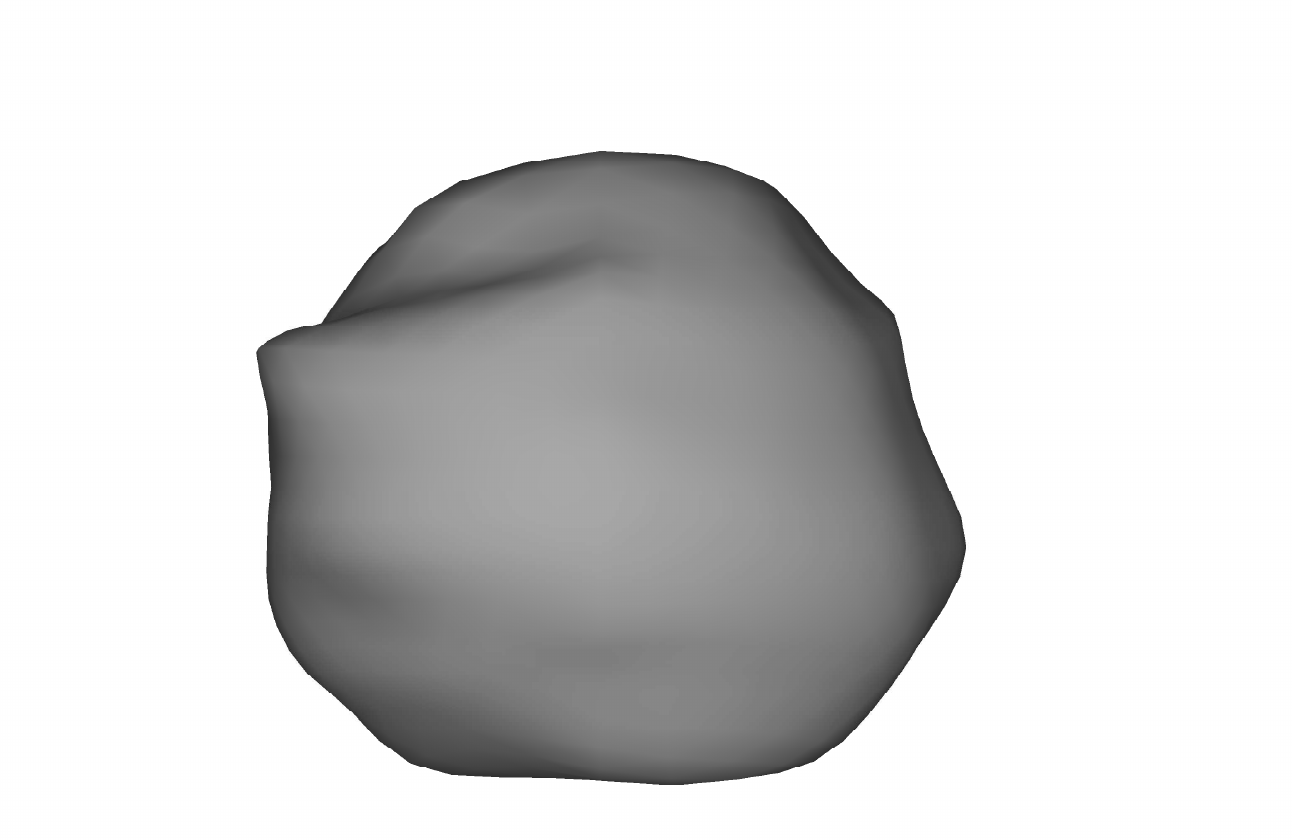}
                \end{subfigure}
  \begin{subfigure}[b]{0.3\columnwidth}
                \includegraphics[clip=true,trim=140 300 100 250,scale=0.25]{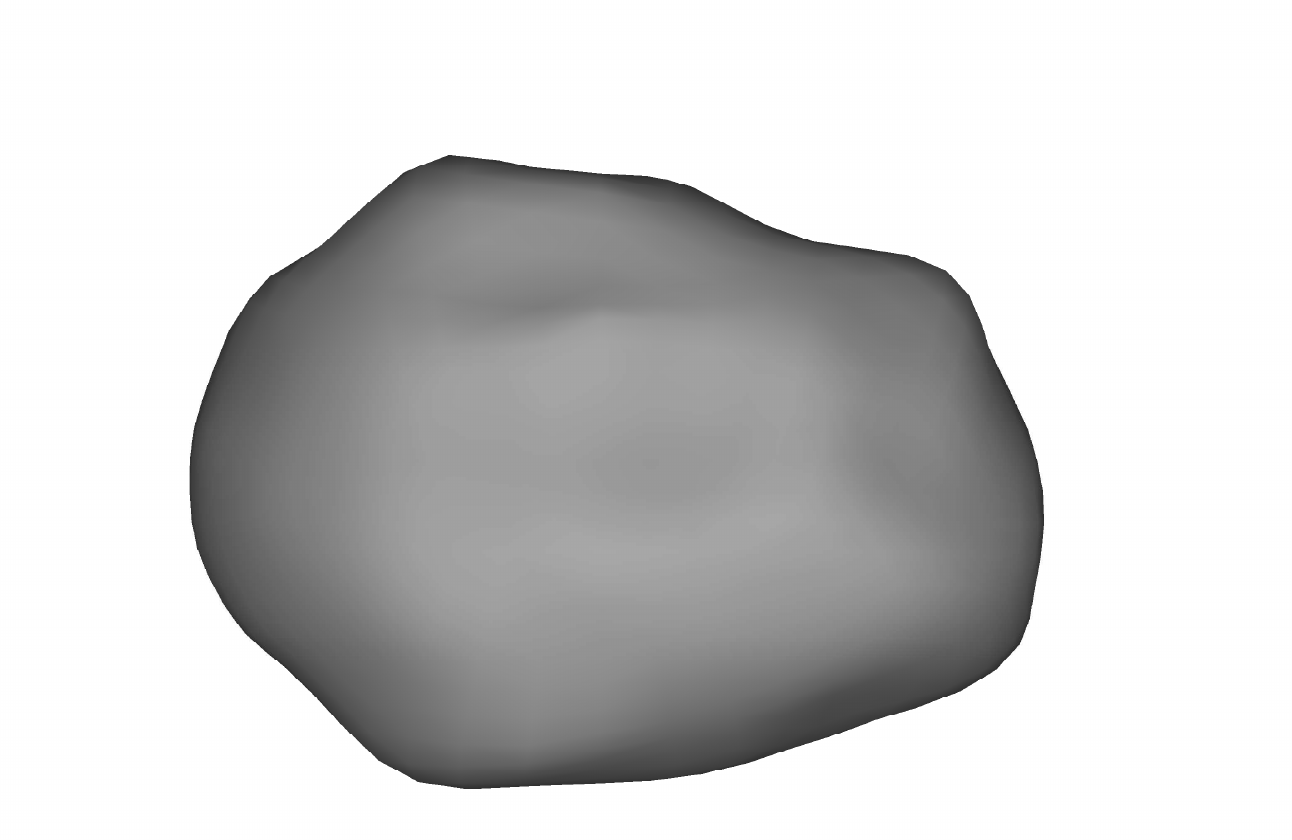}
                \end{subfigure}
   \begin{subfigure}[b]{0.3\columnwidth}
                \includegraphics[clip=true,trim=140 300 100 250,scale=0.25]{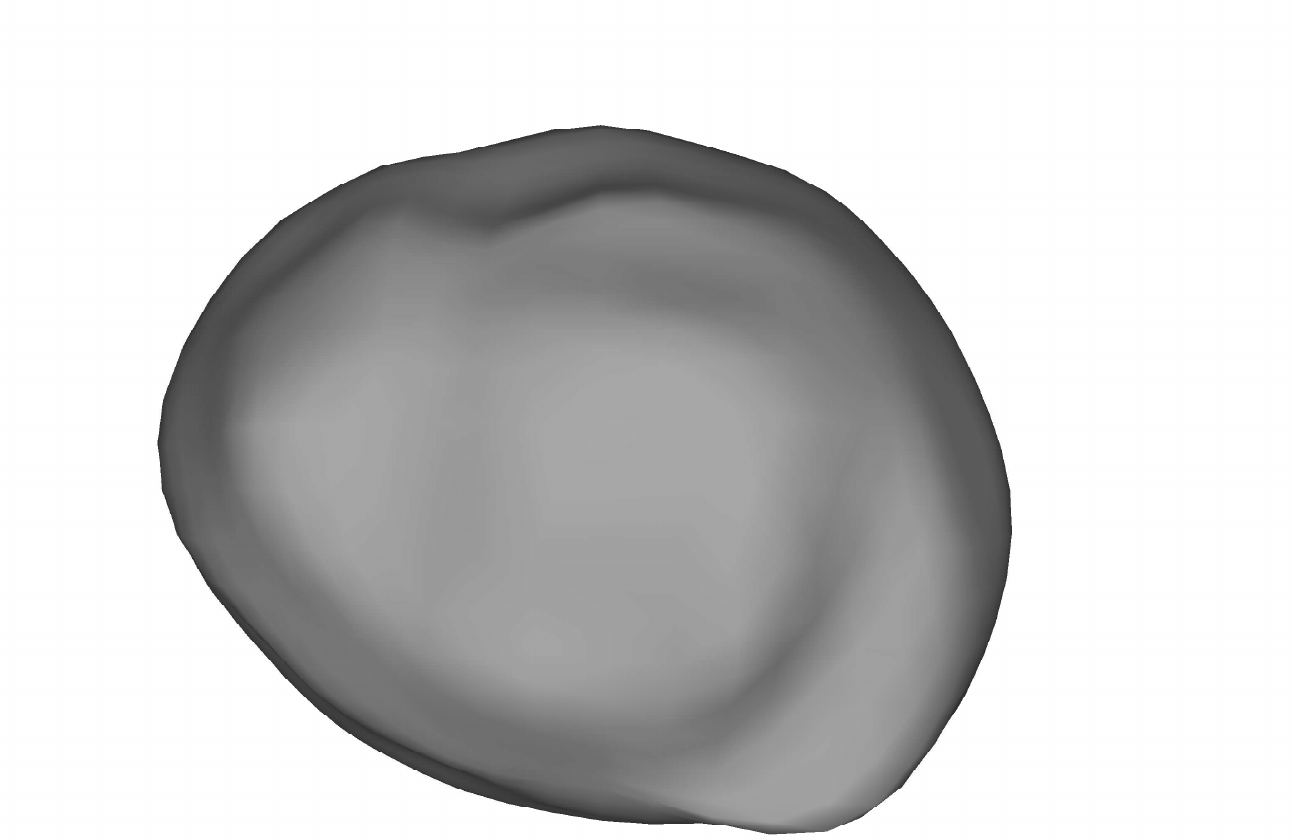}
                \end{subfigure} 
                \caption{\label{fig:ET70}Mid-resolution shape model of the asteroid $2000$ ET${_{70}}$ reconstructed from radar images. Viewing directions are from the positive $x$, $y$, and $z$ axes, respectively.}
                \end{figure}

\begin{figure*}

       \centering

        
         \begin{subfigure}[b]{0.24\textwidth}
                \centering
               \includegraphics[angle=-90,clip=true,trim=70 20 130 20,scale=0.44]{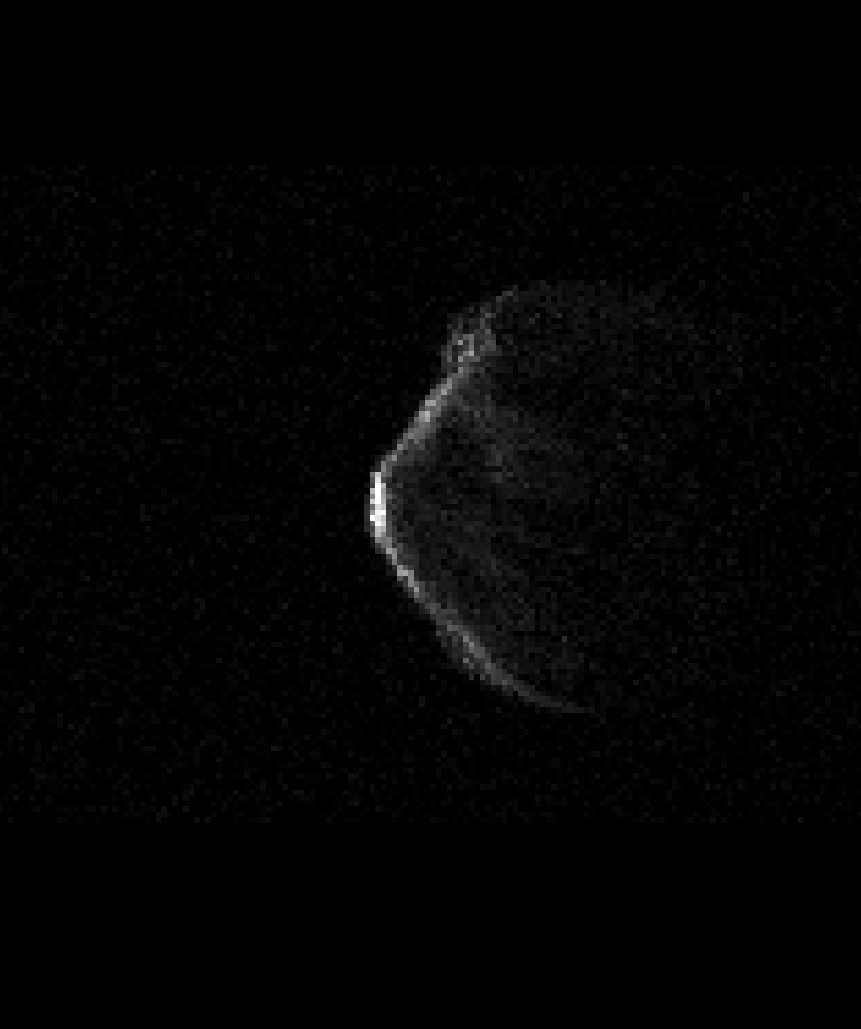}
        \end{subfigure}%
        \begin{subfigure}[b]{0.24\textwidth}
                \centering
                \includegraphics[angle=-90,clip=true,trim=70 20 130 20,scale=0.44]{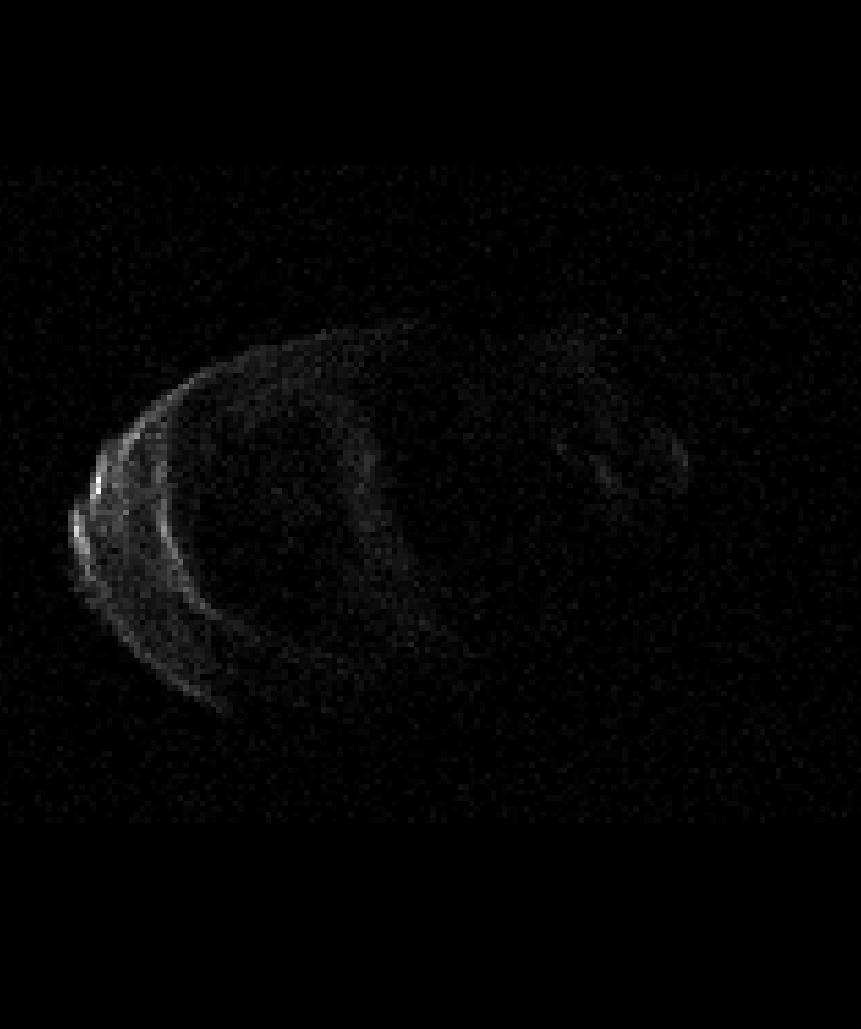}
        \end{subfigure}%
        \begin{subfigure}[b]{0.24\textwidth}
                \centering
                \includegraphics[angle=-90,clip=true,trim=70 20 130 20,scale=0.44]{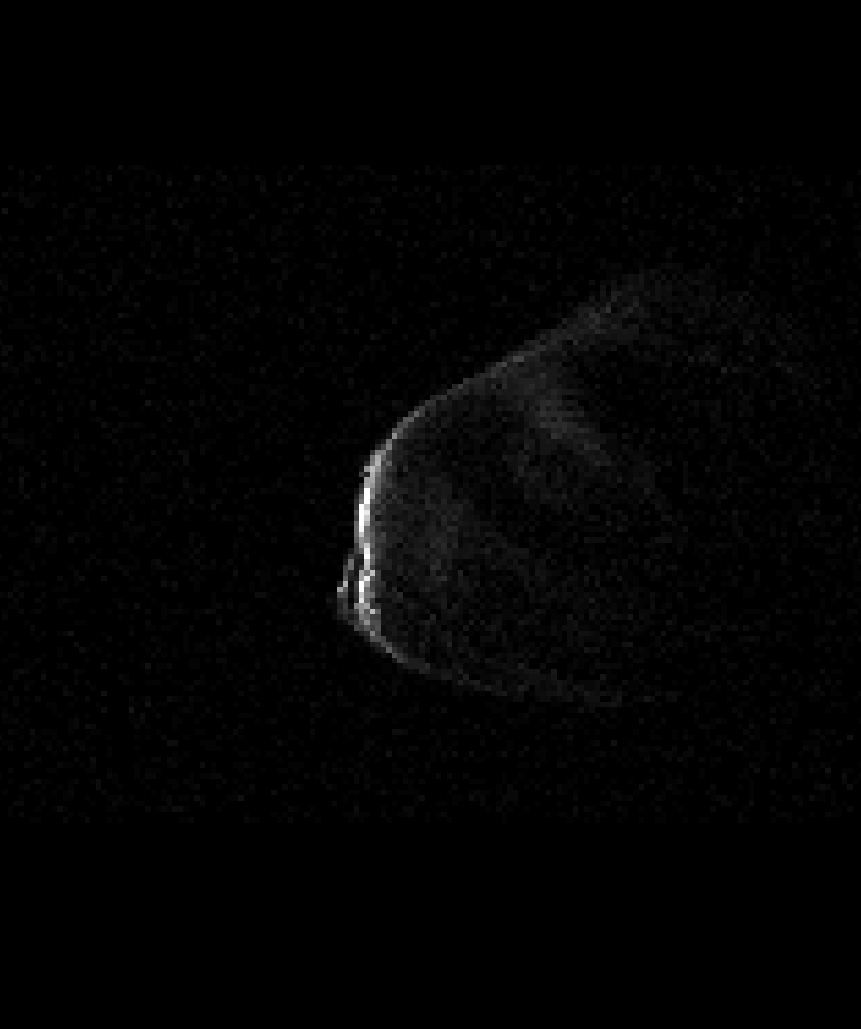}
        \end{subfigure}
        \begin{subfigure}[b]{0.24\textwidth}
                \centering
                \includegraphics[angle=-90,clip=true,trim=70 20 130 20,scale=0.44]{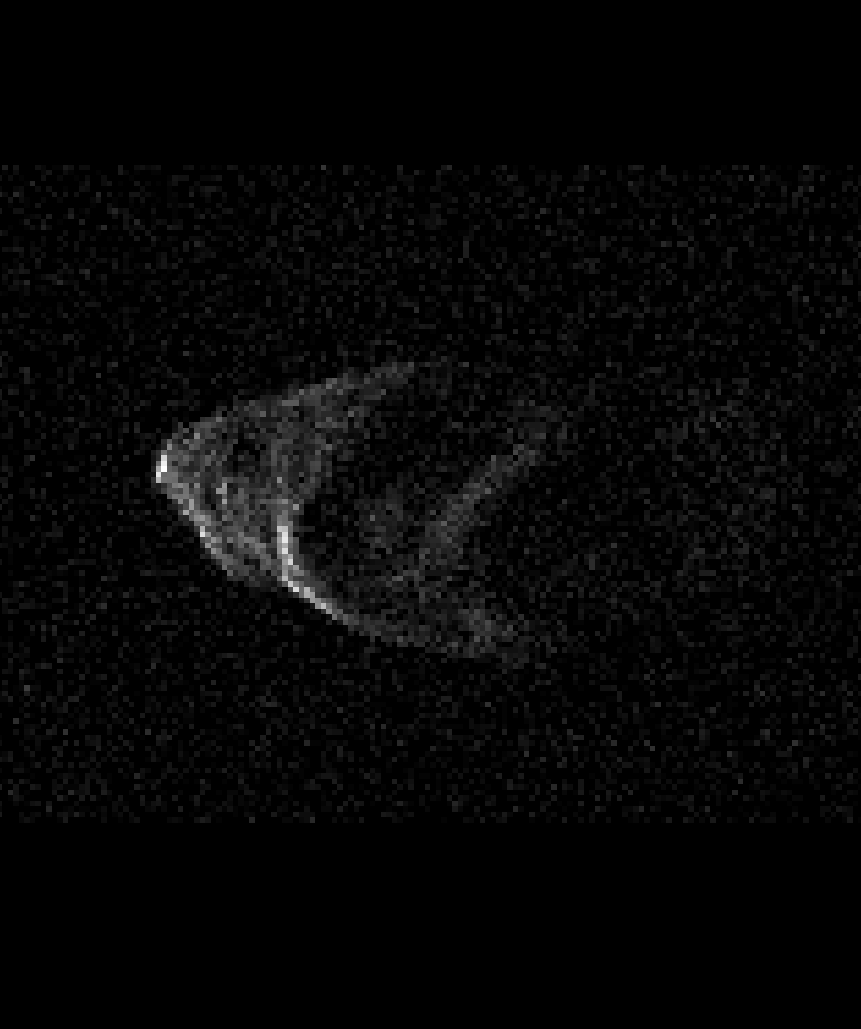}
        \end{subfigure}
          \begin{subfigure}[b]{0.24\textwidth}
                \centering
                \includegraphics[angle=-90,clip=true,trim=70 20 130 20,scale=0.44]{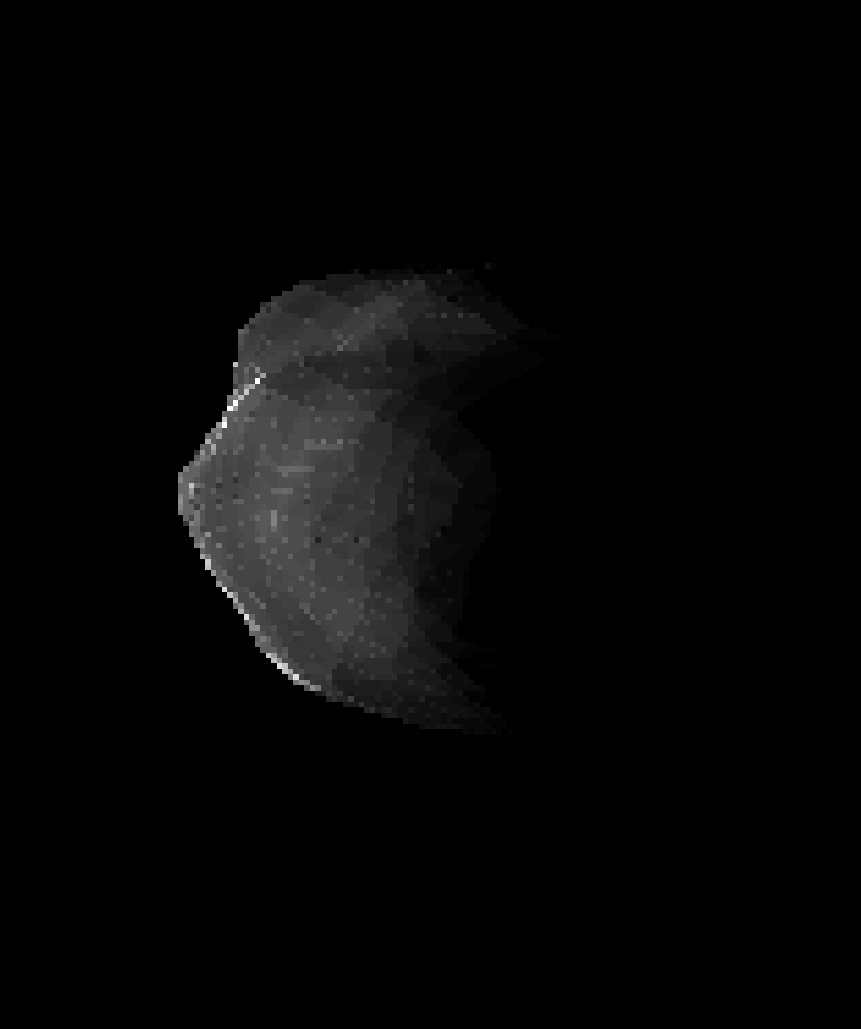}
        \end{subfigure}%
        \begin{subfigure}[b]{0.24\textwidth}
                \centering
                \includegraphics[angle=-90,clip=true,trim=70 20 130 20,scale=0.44]{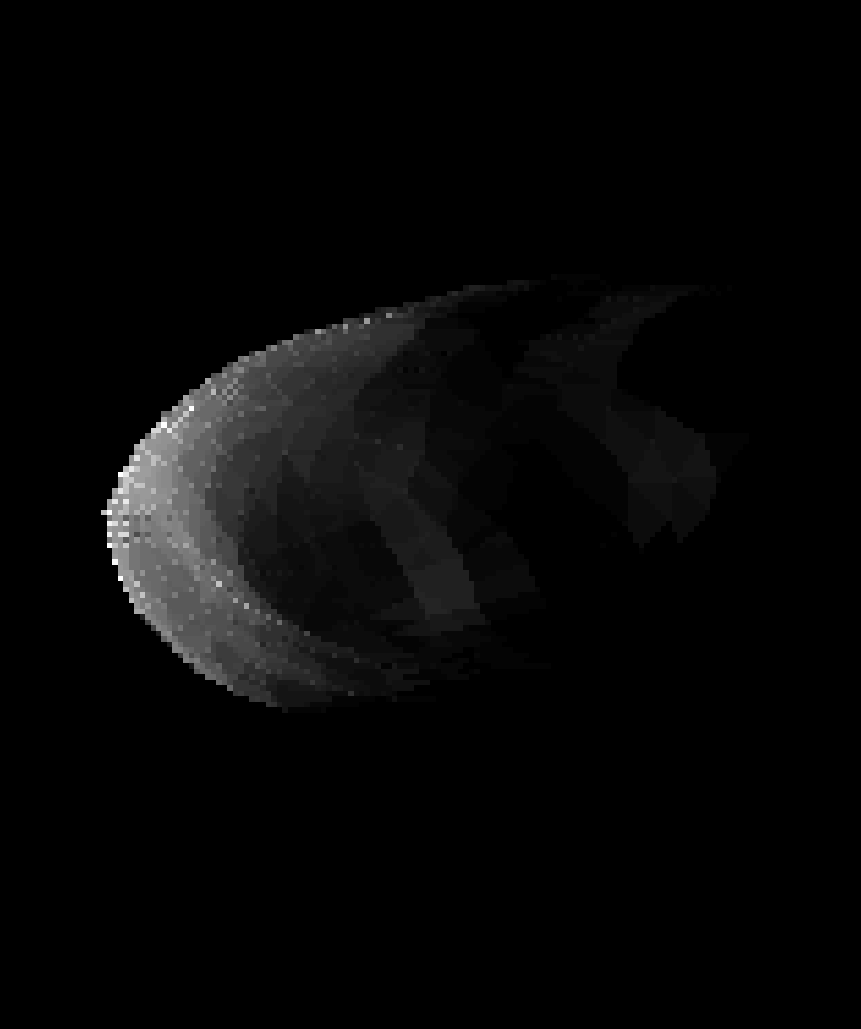}
        \end{subfigure}%
        \begin{subfigure}[b]{0.24\textwidth}
                \centering
                \includegraphics[angle=-90,clip=true,trim=70 20 130 20,scale=0.44]{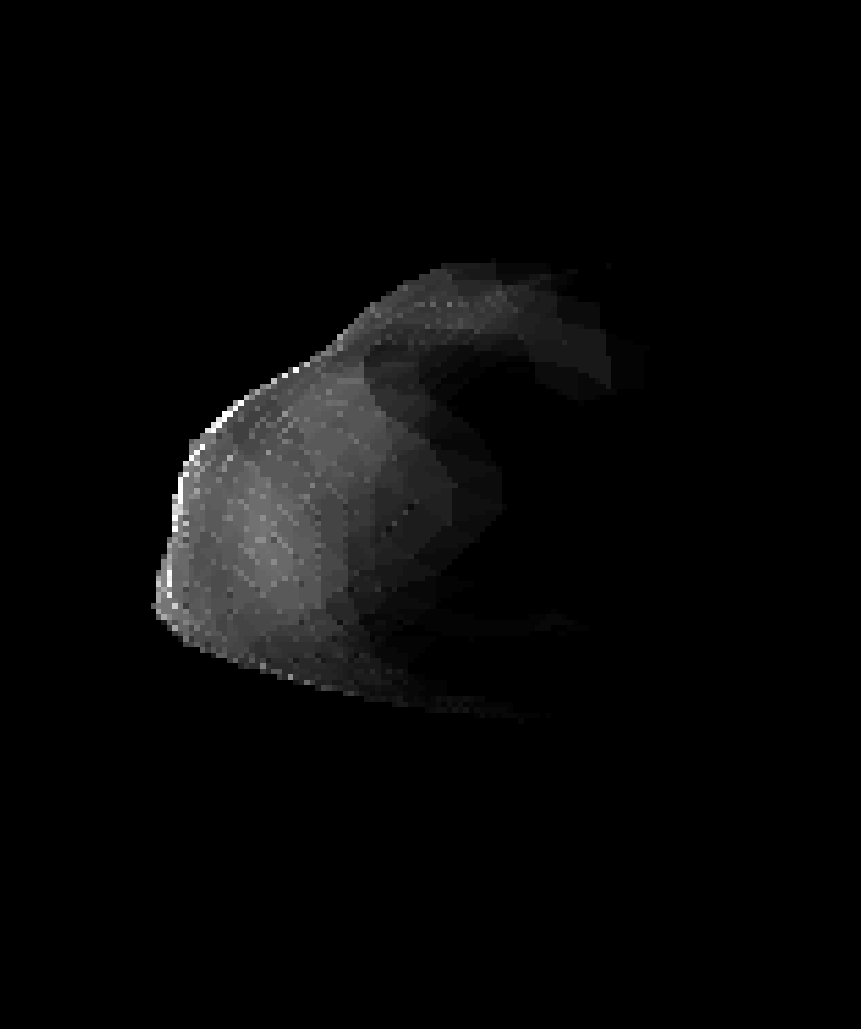}
        \end{subfigure}
        \begin{subfigure}[b]{0.24\textwidth}
                \centering
                \includegraphics[angle=-90,clip=true,trim=70 20 130 20,scale=0.44]{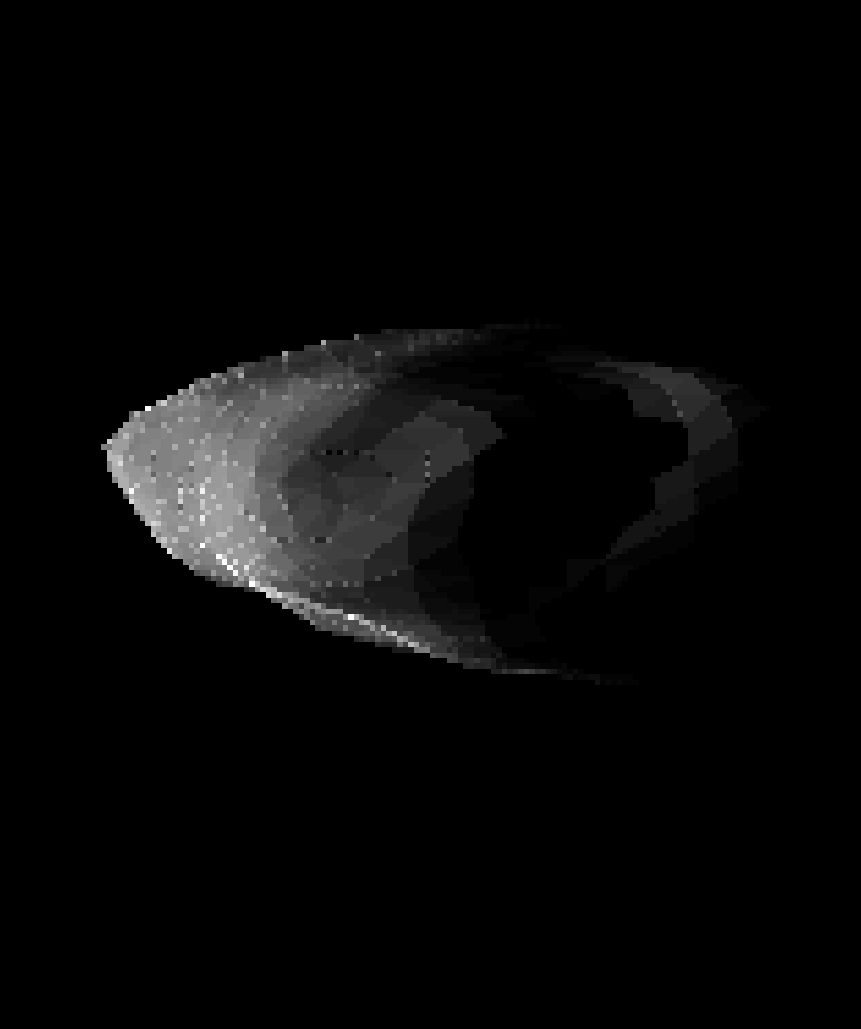}
        \end{subfigure}
        
        \begin{subfigure}[b]{0.24\textwidth}
                \centering
                \includegraphics[angle=-90,clip=true,trim=70 20 130 20,scale=0.44]{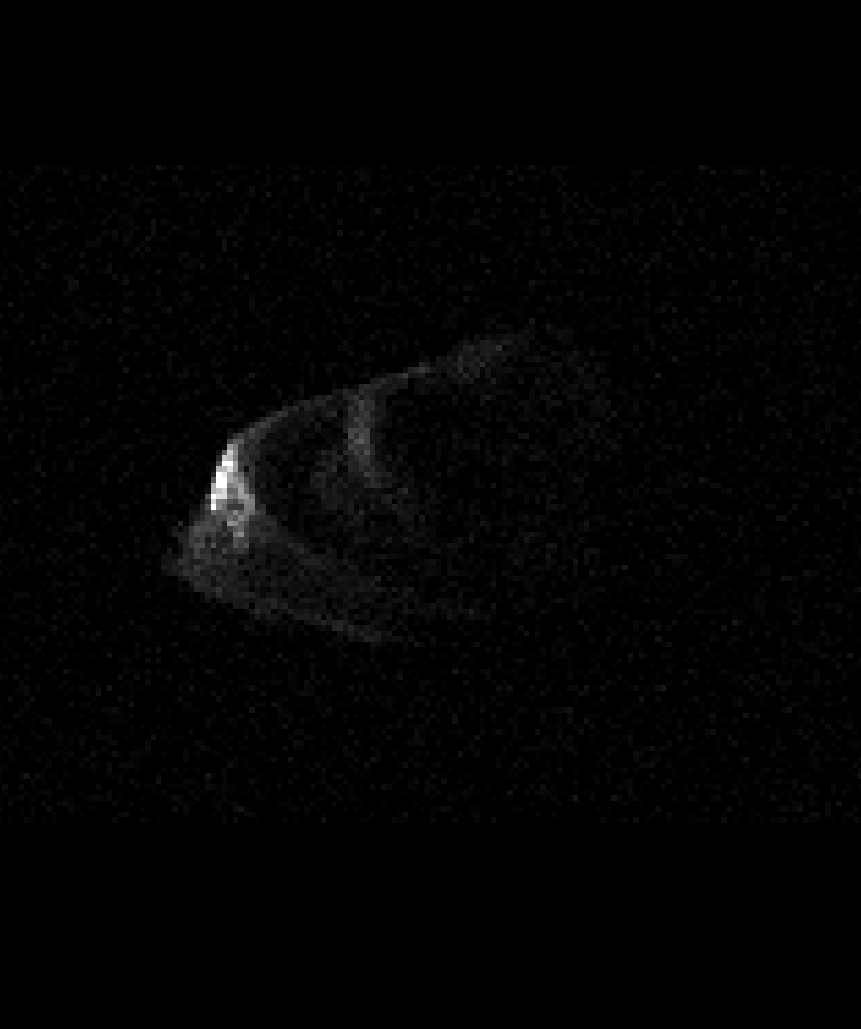}
        \end{subfigure}%
        \begin{subfigure}[b]{0.24\textwidth}
                \centering
                \includegraphics[angle=-90,clip=true,trim=70 20 130 20,scale=0.44]{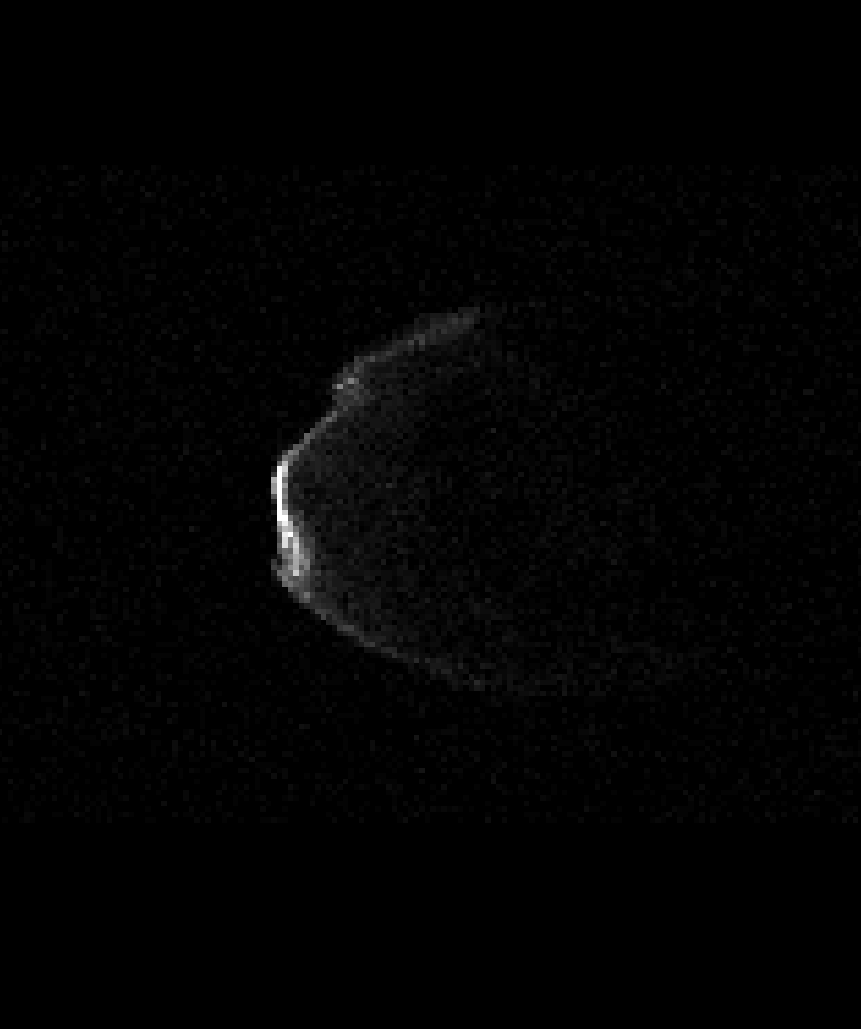}
        \end{subfigure}%
        \begin{subfigure}[b]{0.24\textwidth}
                \centering
                \includegraphics[angle=-90,clip=true,trim=70 20 130 20,scale=0.44]{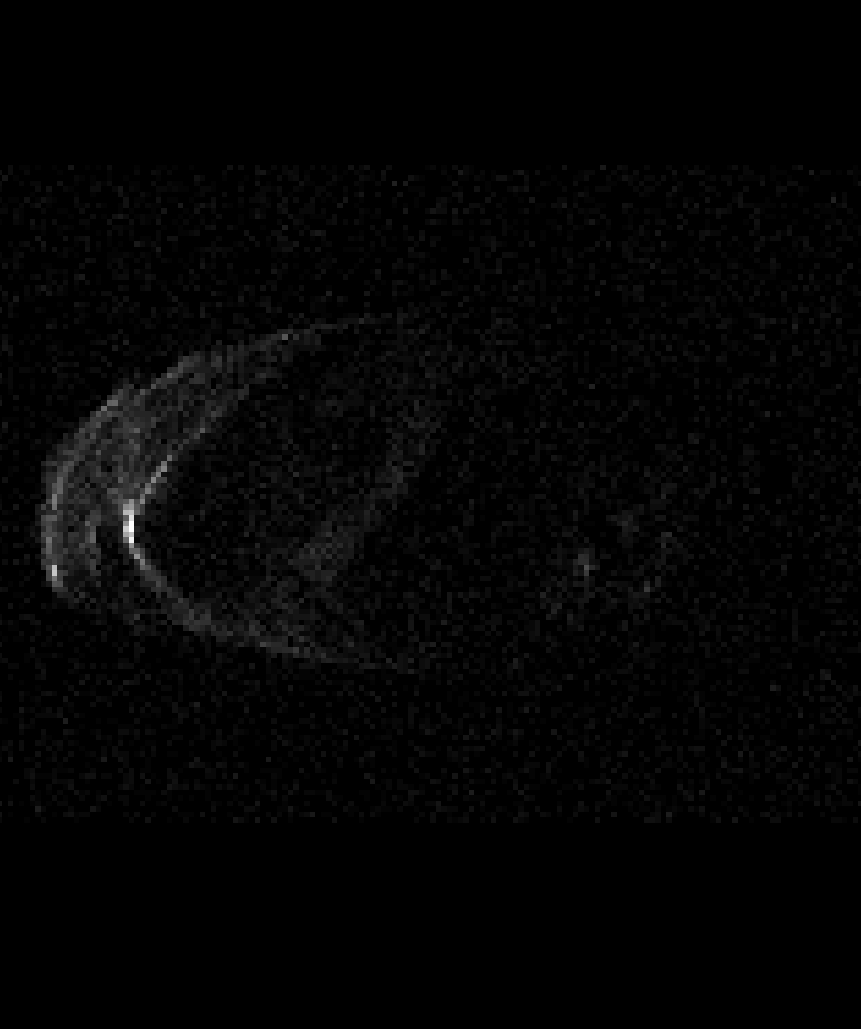}
        \end{subfigure}
        \begin{subfigure}[b]{0.24\textwidth}
                \centering
                \includegraphics[angle=-90,clip=true,trim=70 20 130 20,scale=0.44]{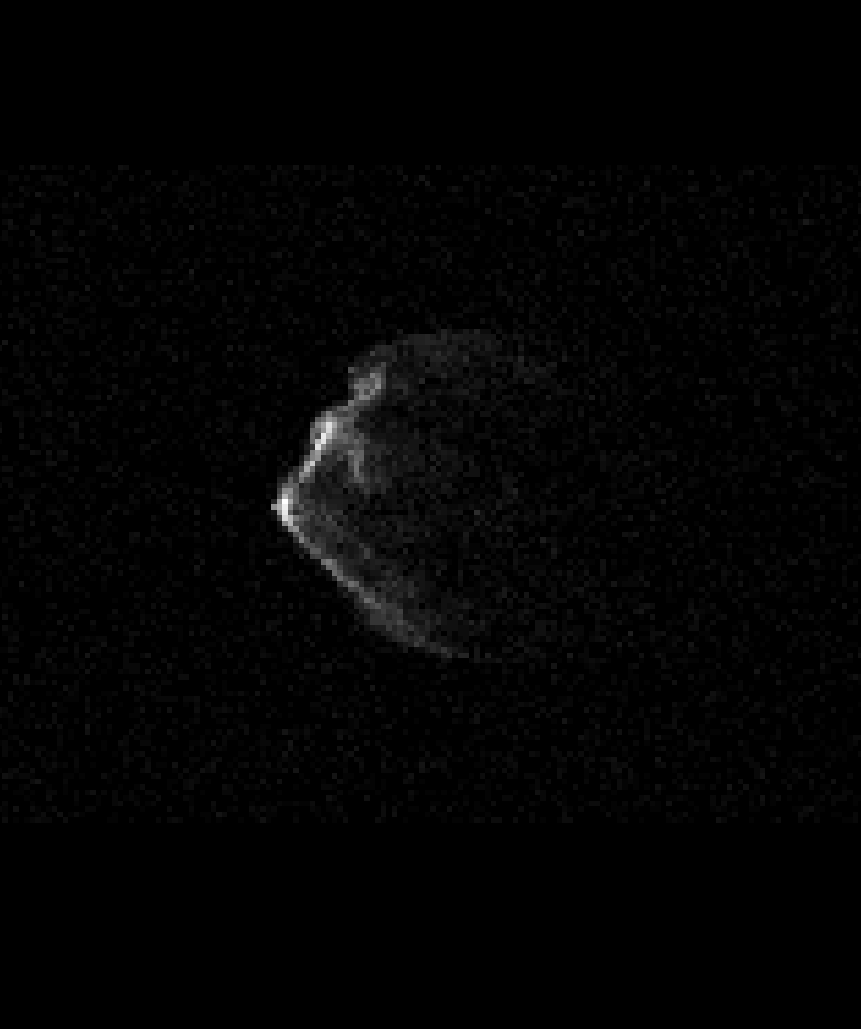}
        \end{subfigure}
         \begin{subfigure}[b]{0.24\textwidth}
                \centering
                \includegraphics[angle=-90,clip=true,trim=70 20 130 20,scale=0.44]{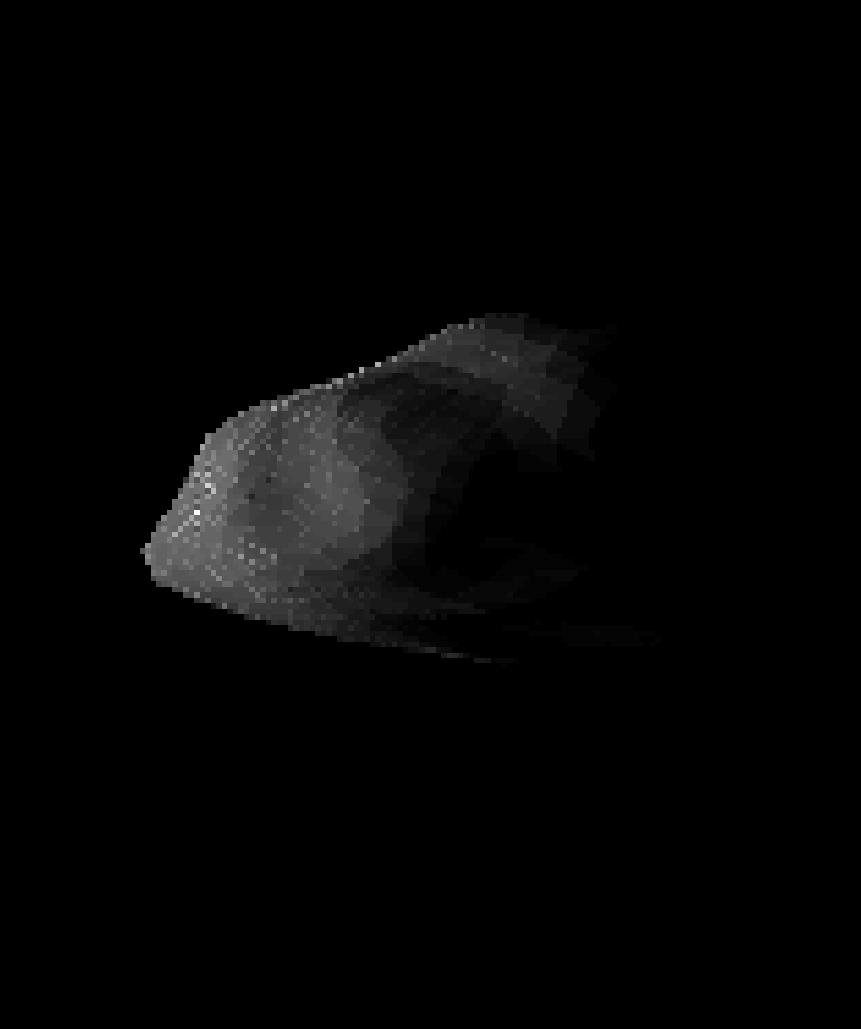}
        \end{subfigure}%
        \begin{subfigure}[b]{0.24\textwidth}
                \centering
                \includegraphics[angle=-90,clip=true,trim=70 20 130 20,scale=0.44]{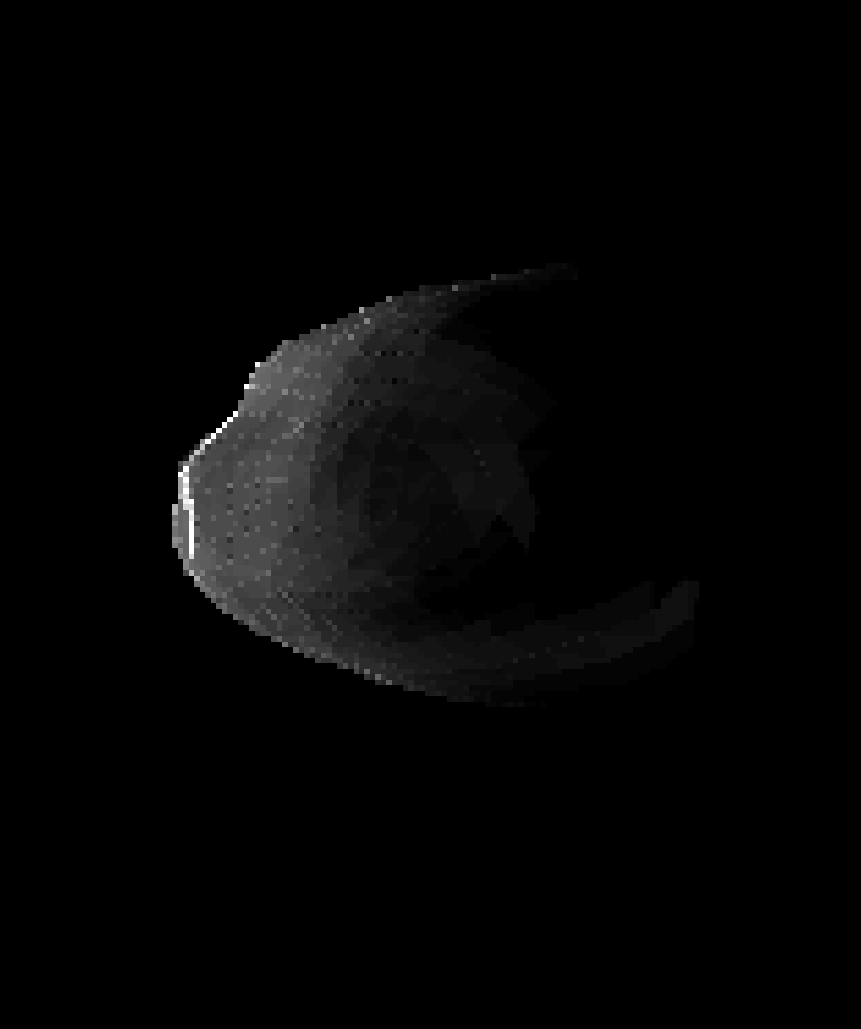}
        \end{subfigure}%
        \begin{subfigure}[b]{0.24\textwidth}
                \centering
                \includegraphics[angle=-90,clip=true,trim=70 20 130 20,scale=0.44]{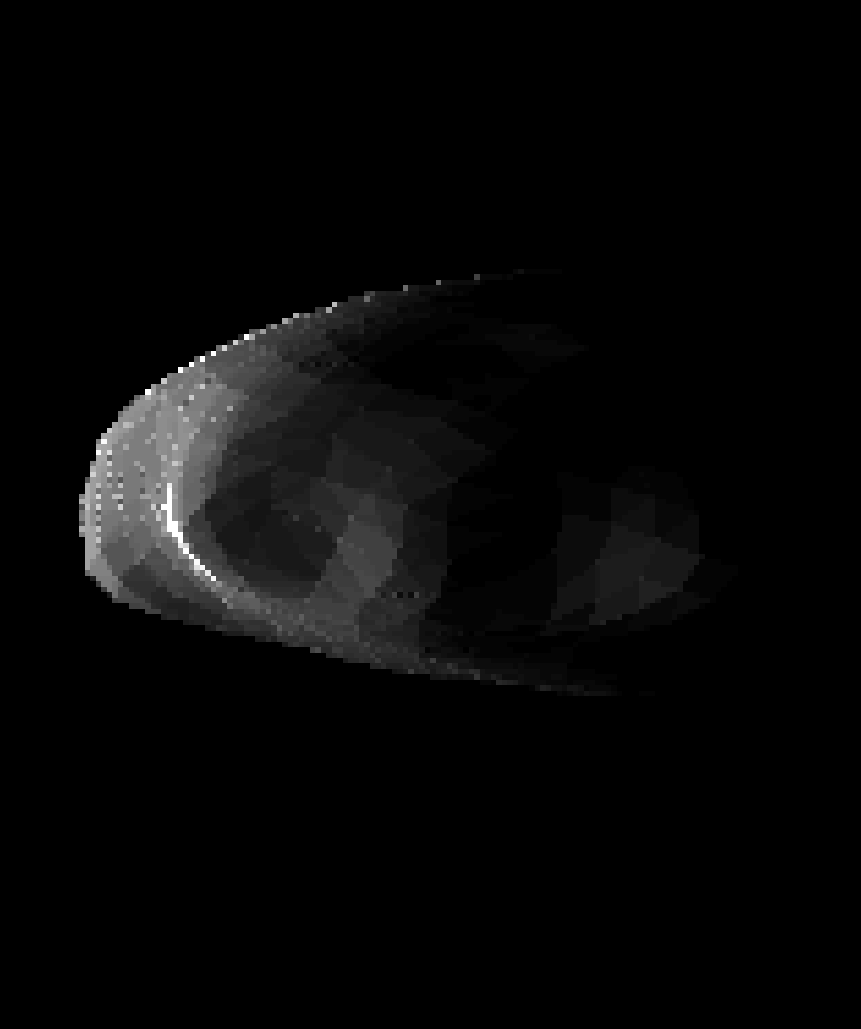}
        \end{subfigure}
          \begin{subfigure}[b]{0.24\textwidth}
                \centering
                \includegraphics[angle=-90,clip=true,trim=70 20 130 20,scale=0.44]{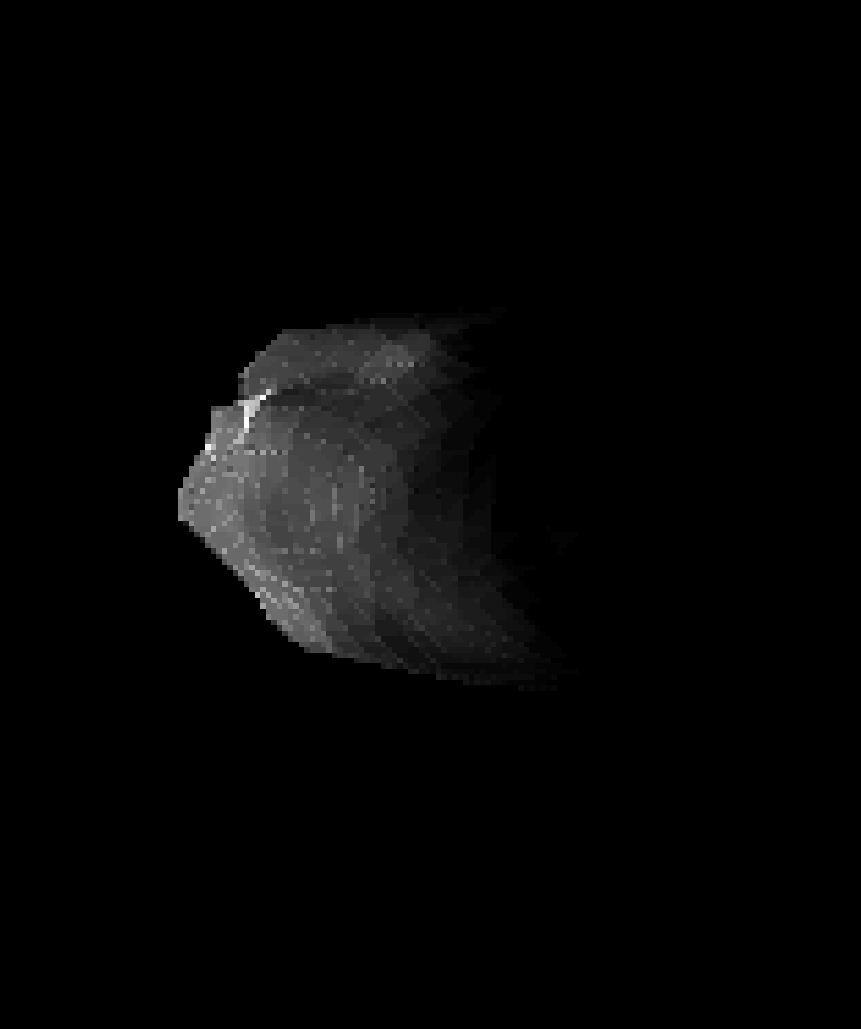}
        \end{subfigure}
       \caption{\label{fig:radar}Examples of range-Doppler images of the asteroid $2000$ ET$_{70}$ from Arecibo  observatory (rows $1$ and $3$) and corresponding simulated  images from the mid-resolution model (rows $2$ and $4$). The contrast scale of the model image is somewhat modified to reveal inner image features.}
\end{figure*}

For each data image, we fit, in addition to the shape parameters, the offset with respect to the model center of mass and the reflectivity term in Eq.\ (\ref{eqcoslaw}). The reconstructed middle-resolution shape is shown in Fig.\ \ref{fig:ET70} and the model fit to the data in Fig.\ \ref{fig:radar}. The shape model fits the boundary contours of the radar images satisfactorily, but there are some differences in the interior details. This is a consequence of the parametrization and  facet size chosen for reconstruction. The interior could be reproduced in greater detail by choosing a different parametrization, for example locally adaptive subdivision surfaces, or by refining the positions of individual vertices. The model dimensions, shape features, and spin parameters agree with those published by \cite{Na13} (the spin parameters are identical except for a $2^\circ$ difference in the pole latitude, well within error limits). 

The main point of the initial low to middle resolution is that the speed of ADAM is considerable, and a detailed knowledge of the instrument or the surface scattering physics is not needed,  so one obtains a first model very fast by just feeding in the images. The middle-resolution radar-based reconstruction (using 82 radar images) was computed in less than an hour on a standard laptop, and GPU programming can reduce the computation time significantly. This makes possible a broad sampling of the parameter space or real-time experimenting with various models. Once a lower-resolution model has been adopted as the final frame, it is  straightforward to refine it further. However, this requires accurate information about the point-spread and scatter functions.

    
 
\subsection{Adaptive optics and other images}
Model reconstruction from adaptive optics images in the Fourier approach is extensively covered in \cite{VK14}, along with an example of the reconstruction of the main belt asteroid Daphne from adaptive optics images and photometry (Fig.\ \ref{fig:daphne}).
Other imaging data may be incorporated into the framework using a similar approach. For instance, flyby images are, from the viewpoint of the reconstruction algorithm, conceptually identical to the AO images. This is one of the attractions of ADAM: at the bare minimum, the user does not need to know anything about the images except their projection matrix and epochs.

We note that the photometric data were actually not even needed in reconstructing Daphne (except for a better period value than with AO images only). The shape results with or without photometry are similar. This shows that even sparse AO data are well sufficient for modelling asteroid spin states and shapes in detail. 

\subsection{One-dimensional projection operators}
In the regime between disk-integrated and disk-resolved observations there are one-dimensional operators that project the plane-of-sky onto a line. Typical examples are the continuous-wave (CW) Doppler spectra that measure the distribution of the reflected power in frequency only, and the fine guidance sensors (FGS) onboard the Hubble Space Telescope, measuring the brightness distribution along an instrument axis. One-dimensional projections are seldom sufficient for actual shape reconstruction, but they may contain useful information about the object's size or indications about the bifurcated structure \citep{KV12}, and combined with other sources they facilitate shape inversion. 

In both examples, the measurement can be written in the form
\begin{equation}
\label{ope1} S(x)=\int I(\xi,\eta)\,P(x-\xi\cos\gamma-\eta\sin\gamma)\,\diff\xi \,\diff\eta,
\end{equation}
where $I(\xi,\eta)$ is the plane-of-sky brightness (optical or radar) distribution of an object, $P$ is the point-spread function of the instrument, and the angle $\gamma$ corresponds to the rotation of the sensor in the image plane.
In \cite{KV12}, the integral was evaluated using a Monte Carlo method: the projected model was sprinkled with uniformly distributed sampling points, and the integral was approximated as a sum over the visible and illuminated sampling points. Here we demonstrate how the Fourier transform method can be used to interpret the integral as a tomographic operator on the Fourier plane.

Taking the Fourier transform on both sides and using the projection-slice theorem  \citep[a slice of a 2-D FT along a line through the origin equals the 1-D FT of the projection of the original 2-D function onto a line in the same direction; see e.g.][]{Br04}, we get
\begin{equation}
 \mathcal{S}(f)=\mathcal{I}(f\cos\gamma,-f\sin\gamma)\,\mathcal{P}(f),
\end{equation} 
where the calligraphic characters denote the Fourier-transformed functions. Now it is obvious that $\mathcal{S}(f)$ is a slice of a 
Fourier-transformed brightness distribution  along a line through the origin, multiplied with the Fourier transform of the point-spread function. Moreover, this means that the same algorithm may be used to fit both FGS and adaptive optics data, and similarly both CW Doppler data and the range-Doppler images. In other words, we extract a one-dimensional Fourier transform from the 2-D model FT, and compare this with the 1-D FT formed from the data in the same way as in the full 2-D case.

\section{ADAM algorithm}
The flowchart in Fig.\ \ref{fig:adam} describes the workings of ADAM. More specifically, the algorithm may be divided in five distinct steps:
\begin{enumerate}
\item For each data image $D_i$ and observation geometry $\mathcal{E}_i$, the two-dimensional Fourier transform $\mathcal{F}D_i(u,v)$ of $D_i$ is sampled at a set of points $\{(u_{ij},v_{ij})\}$, $j=1\ldots N_i$, on the spatial frequency plane. The size of the set is chosen to correspond to the level of resolution. For pixel images, the transform can be computed by Eq.\ (\ref{foutrans}) when considering each pixel as a polygon, or by using fast Fourier transform functions for chosen grid points (but the time spent for  $\mathcal{F}D_i(u,v)$ is irrelevant as most of the computations are for the trial models). 
\item The shape support and resolution level (number of parameters) are chosen. The parameters are initialized such that the initial shape is a sphere approximately equal in size to the target.
\item For each observation geometry $\mathcal{E}_i$, the Fourier transform $\mathcal{F}M_i(u,v)$ of the corresponding projection image $M_i$ of the model is calculated as described in the previous sections, together with the partial derivatives of $\mathcal{F}M_i(u,v)$ with respect to all optimized parameters. Ray-tracing, scattering or luminosity models, and coordinate transforms for the image plane are discussed in \cite{KT01b}, \cite{Ka11}, and \cite{VK14}.
\item An objective function $\chi^2$ is formed, with the square norm of the complex-valued FT fit error:
\begin{equation}
\begin{split}
&\sum_i\sum_{j=1}^{N_i}\left\Vert\mathcal{F}D_i(u_{ij},v_{ij})-e^{2\pi\imath\left(o^x_i u_{ij}+o^y_i v_{ij}\right)}\,\mathcal{S}_i(u_{ij},v_{ij})\,\mathcal{F}M_i(u_{ij},v_{ij})\right\Vert^2\\&+\sum_i\lambda_i\gamma_i^2=:\chi^2,\end{split}\label{chisq}
\end{equation}
where $(o^x_i,o^y_i)$ is the offset between the data image $D_i$ and the model image $M_i$, and, by the convolution theorem, $\mathcal{S}_i$ is the Fourier transform of the point-spread function of the imaging system. The  $\gamma_i$ represent various regularization terms defined above.

For brevity, we have written only one data mode in Eq.\ (\ref{chisq}); any number of modes with their goodness-of-fit functions can be added to the sum. These functions for photometry and silhouettes (occultations) are given in \cite{KT01b}, \cite{Ka11}, and \cite{VK14}. The determination of the weights of the data modes (as $\lambda_i$ for the regularization functions) is discussed in \cite{Ka11} and \cite{KV12}. Weights can be determined for any subsets of data (e.g., less reliable images) if necessary.

In addition, the intensity level of each data and model image must be normalized. Often it is enough to divide both model $M_i$ and data image $D_i$ by their respective mean intensities. Equivalently, writing
\begin{equation}
\chi^2:=\sum_{ij} \left\Vert\mathcal{D}_i(u_{ij},v_{ij})-\mathcal{\tilde M}_i(u_{ij},v_{ij}) \right\Vert^2+\lambda\gamma^2,
\end{equation}
we have 
\begin{equation}
\chi^2_{\rm rel}=\sum_{ij} \left\Vert\frac{\mathcal{D}_i(u_{ij},v_{ij})}{\langle{\mathcal\Vert D}_i\Vert\rangle}-\frac{\mathcal{\tilde M}_i(u_{ij},v_{ij})}{\langle{\mathcal\Vert {\tilde M}}_i\Vert\rangle} \right\Vert^2+\lambda\gamma^2,
\end{equation}
where the mean $\langle \cdot\rangle$ is taken over $\{(u_{ij},v_{ij})\}$, $j=1\ldots N_i$. However,
sometimes it is better to allow the intensity level of each $M_i$ to be a free parameter and use $\chi^2$; this is useful in the case where the mean intensity of $D_i$ is corrupted by excessive noise in the image background (this is typical for range-Doppler images). This causes the $\chi^2_{\rm rel}$-based solution to have a slightly wrong size to compensate for the ``diluted'' normalized intensity level inside the actual object region of $D_i$. 

\item The shape and spin parameters and the offsets $(o^x_i,o^y_i)$ as well as the possible intensity level factors $C_i$ minimizing $\chi^2$ are determined with a suitable method such as the Levenberg-Marquardt algorithm. If there are several hundreds of parameters, as in the case of fitting all shape vertices directly (instead of using function series or control points) to produce maximal resolution, the conjugate gradient method is efficient \citep{KT01b}.

\end{enumerate} 
\tikzstyle{decision} = [diamond, draw, fill=blue!20, 
    text width=4.5em, text badly centered, node distance=2cm, inner sep=0pt,font=\tiny]
\tikzstyle{block} = [rectangle, draw, fill=blue!20, 
    text width=4.5em, text centered, rounded corners,node distance=1.5cm, font=\tiny]
\tikzstyle{line} = [draw, -latex']
\tikzstyle{cloud} = [draw, ellipse,fill=red!20, node distance=3cm,
    minimum height=2em]
\begin{figure}

\begin{tikzpicture}[node distance = 1cm, auto]
    \node [block] (para) {Shape parameters};
   
    \node [block, below of=para] (trimesh) {Surface mesh};
    \node [block, below of=trimesh] (camera) {Transform to camera frame};
    \node [block, left of=camera, node distance=5cm] (obs) {Observation};
    \node [block, below of=camera] (proj) {2D projection};
    \node [block, below of=proj] (FT) {Fourier transform};
  \node[block, left of=FT,node distance=2.5cm](OTF){Instrument PSF};
    \node[decision,below of=FT] (opt){Optimization};
    \node[block,left of=opt, node distance=5cm](data) {Data};
    \node[block,left of=opt,node distance=2.5cm](FFT) {FFT};
    \node[block,below of=opt](reg) {Regularization};
 
    \path [line] (obs) -- (camera);
    \path [line] (para) -- (trimesh);
    \path [line] (trimesh) -- (camera);
    \path [line] (camera) -- (proj);
    \path [line] (proj) -- (FT);
    \path [line] (FT) -- (opt);
    \path [line] (FFT) -- (opt);
    \path [line] (data) --(FFT);
    \path[line] (obs) -- (data);
    \path[line] (reg) -- (opt);
    \path[line](OTF) --(FT);
    \path[line] (opt) node[yshift=5em,xshift=3.7em,rotate=90,font=\tiny] {update parameters} -|($(para.east)+(0.7,0)$)|- (para);
\end{tikzpicture}
\caption{\label{fig:adam} ADAM optimization algorithm as a schematic for one image type.}
\end{figure}
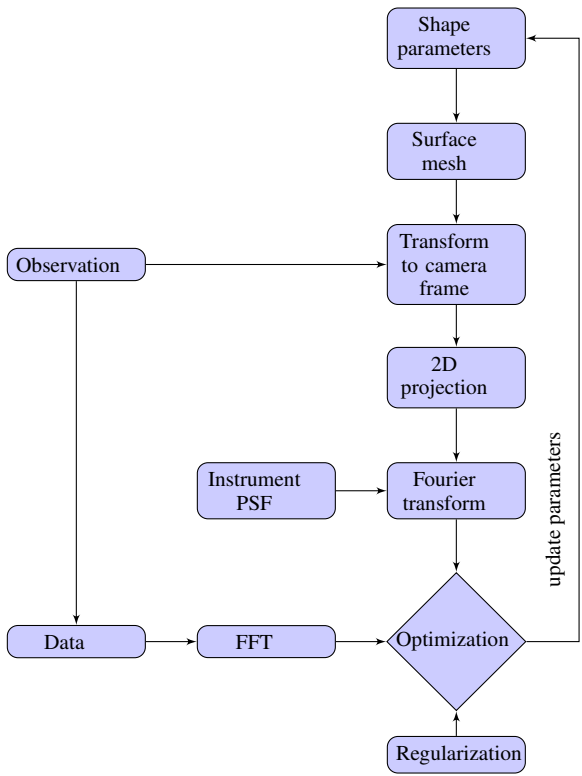

\section{Conclusions and discussion}

ADAM can handle radar data, images, interferometry (also in the thermal infrared), photometry, and occultations separately or in combinations.
The ADAM procedure consists of a number of modules, and there are various options for each module, customized to the end-user (e.g., the adopted optimization method, regularization functions, shape support and mesh structure, ray-tracing method, coordinate system, luminosity/scatter model, image formats, etc.). In this sense, ADAM is a toolbox and a set of building blocks rather than a ready-made program.

The main idea behind ADAM is the efficient use of the Fourier transform in handling both images and one-dimensional projection data. Fourier analysis has long been used in, e.g., image compression because it captures the essential information conveniently in a hierarchy of resolution. In the same vein, the FT approach in ADAM is ideal for producing models of desired levels of resolution, especially in the low- to medium-resolution category. In this framework, the goodness-of-fit function between the model and the data is easy to compute and use in optimization. What is more, its convergence properties are more robust than if the images were used directly (in fact, one does not necessarily even have to look at the images or know much about the instrument that produced them). An analogy of this paradox is the simple one-dimensional problem of realigning two phase-shifted copies of a dual-frequency signal. If one does this by minimizing the signal difference by optimizing the shift in the original amplitude space, there are multiple local minima, but in frequency space the offset is found immediately.

Despite its automatic character, ADAM should not be used as a black box: asteroid reconstruction is a complicated inverse problem, and one should be familiar with its mathematical principles to understand the limitations and information content of the data sources.

\begin{acknowledgements}
This work was supported by the Academy of Finland project
``Modelling and applications of stochastic and regular surfaces in
inverse problems'' and the CoE in inverse problems research. The work of J\v{D} was supported by the grant GACR P209/10/0537 of the Czech Science Foundation.
We thank Jean-Luc Margot for providing the radar images of asteroid 2000 ET$_{70}$.

\end{acknowledgements}

\bibliographystyle{aa}
\bibliography{ADAM}

\section*{Appendix: Sample ADAM functions}

In order to make the structure of ADAM more concrete, we show an example of how the program is divided into subroutines. We consider only the part of the program which computes the heat flux density of an object and its Jacobian (i.e., the partial derivatives of each modelled flux data point w.r.t. the free parameters), as all the other modules are structurally similar. In the case of optical images, the flux is replaced by brightness from scattering; for radar, by the signal strength in the range-Doppler plane.

The partial derivatives w.r.t. the shape parameters are initially calculated with respect to the vertex coordinates, making the routines independent of the parametrization used. Only in the final phase is the Jacobian determined using the chain rule. The functions are complex-valued, since the fitting is done on the frequency plane. In the optimization, the data are divided into real and imaginary parts and fitted separately. Usually, the Levenberg-Marquardt or the conjugate gradient method is used to optimize the $\chi^2$-fit.

ADAM uses three different coordinates systems: the asteroid-centric coordinate frame with coordinate axes fixed to the asteroid, the asteroid-centric inertial frame, and the camera frame, which is determined by the instrumental orientation geometry. The plane-of-sky view of an asteroid is obtained by projecting the asteroid in the camera frame to the xy-plane.

The Jacobian and the vector consisting of simulated values corresponding to the observations are computed using the following subroutines:
\begin{itemize}
\item \texttt{Generate\_HF\_Matrix} calls the subroutine \texttt{Calc\_Heat\_Flux} for each observation, and then combines the Jacobian submatrices into a full Jacobian matrix corresponding to all the observations. 
\item \texttt{Calc\_Heat\_Flux} Calculates the Fourier transform of the two-dimensional plane-of-sky  flux density and its partial derivatives by calling the subroutines \texttt{Calc\_Temp}, \texttt{Rot\_Matrix}, \texttt{Cam\_Matrix}, \texttt{Calc\_Vis} and \texttt{Calc\_FT}. After the flux density of each facet is determined, the routine transforms the triangular mesh to the camera frame and projects the visible part of the mesh  onto the xy-plane by discarding the z-coordinate. Finally, each vertex is transformed to the frequency plane using the \texttt{Calc\_FT} subroutine and the contributions of the Fourier-transformed facets are summed.
\item \texttt{Calc\_Temp} determines the temperature of the facets corresponding to the observation geometry, using the FFT method. The partial derivatives of the temperature with respect to the shape parameters are also calculated. This subroutine also calls subroutines \texttt{Rot\_Matrix} and \texttt{Calc\_Vis}.
\item \texttt{Rot\_Matrix} calculates the rotation matrix needed to transform the object to the inertial frame. The rotation matrix is determined by the spin vector and the observation time.
\item \texttt{Cam\_Matrix} determines the matrix needed to transform the inertial frame to the camera frame. This depends on the instrument location and orientation. The z-coordinate codes the  relative distance from the instrument.
\item \texttt{Calc\_Vis} determines the visibility of facets using ray-tracing. In contrast to the optical case, a facet can be visible to the observer even if it is not illuminated by the sun.  
\item \texttt{Calc\_FT} Calculates the Fourier transform of a triangle projected onto the xy-plane together with the corresponding partial derivatives.
\end{itemize}

The most important setup factors determining the computation time of ADAM are the numbers of facets and data points (the number of images and their pixels). The computation time increases approximately linearly with both numbers. The cost of actual optimization steps increases superlinearly with the number of free parameters, but with large datasets (such as the radar example above) most of the computation is spent on determining function values and their partial derivatives with respect to the vertex coordinates. In such cases, the number of shape parameters is not critical for the computational cost in the mid-resolution regime, so one is free to choose a number that best corresponds to the resolution level (and set the number of facets accordingly).  When the dataset is small, the computation time is short in any case, and the model is likely to be a low-resolution one, so again the number of parameters is not an issue.
The cost of visibility determination by ray-tracing is insignificant as the potential blocker facets can be precomputed \citep{KT01a}. 

The shape reconstruction from observations is an easily parallelizable problem. There are two obvious levels of parallelism: each observation can be calculated independently; or, within each observation, the contribution of each facet may be determined simultaneously. The best choice depends on the computer's architecture. The observation-level parallelism may be easily exploited using the MATLAB parallel computing toolbox, or more effectively by using the OpenMP API in the C language. The reduction in execution time scales almost linearly with the number of CPU cores. This is the approach currently implemented in ADAM.

While it is possible to implement facet-level parallelism on the CPU by dividing the facet computations between several CPU cores, a more natural approach is to use one thread per facet. This kind of  implementation  is inefficient on the CPU, since the thread-switching latency is high compared to the running time of a thread. However, the ability of the GPU to run thousands of lightweight threads simultaneously combined with the virtually costless thread switching makes it possible to attain orders of magnitude faster computation than with CPU. GPU-accelerated computing will be implemented in ADAM using the Nvidia CUDA programming platform.

\end{document}